\documentclass{iopart}
\usepackage{epsfig,newlfont}
\usepackage{subfigure} 
\DeclareMathAlphabet{\mathitbf}{T1}{cmr}{bx}{it}
\usepackage[colorlinks, 
linkcolor=blue,
urlcolor=blue,
citecolor=blue,
bookmarksopen=false]{hyperref}

\begin{document}
 
\title{Study of the neoclassical radial electric field of the TJ-II flexible heliac}

\author{J.L. Velasco$^1$ and F. Castej\'on$^1$}

\address{$^1$ Laboratorio Nacional de Fusi\'on, Asociaci\'on EURATOM-CIEMAT, Madrid, Spain}

\ead{joseluis.velasco@ciemat.es}

\begin{abstract}

%Calculations of the monoenergetic radial diffusion coefficients are
%presented for several configurations of the TJ-II stellarator usually
%explored in operation. The neoclassical radial fluxes and the
%ambipolar electric field for the standard configuration are then
%studied for three different collisionality regimes. Monte Carlo error
%estimation allows us to provide precise results.

Calculations of the monoenergetic radial diffusion coefficients are
presented for several configurations of the TJ-II stellarator usually
explored in operation. The neoclassical radial fluxes and the
ambipolar electric field for the standard configuration are then
studied for three different collisionality regimes, obtaining precise
results in all cases.

\end{abstract}

%Uncomment for PACS numbers title message
%\pacs{00.00, 20.00, 42.10}
% Keywords required only for MST, PB, PMB, PM, JOA, JOB?
%\vspace{2pc}
%\noindent{\it Keywords}: Article preparation, IOP journals
% Uncomment for Submitted to journal title message
%\submitto{\JPA}
% Comment out if separate title page not required
%\maketitle

\section{Introduction}\label{SEC_INTRO}

Radial electric fields are recognized to play a key role in the radial
transport of stellarators. From the neoclassical transport point of
view, they affect the particle
orbits~\cite{helander2002collisional,wakatani1998stellarator}: for
low-collisionality plasmas, they suppress the unfavorable $1/\nu$
regime~\cite{galeev1979theory} and allow for the formation of electron
transport barriers, see Ref.~\cite{yokoyama2007cerc} and references
therein. Additionally, radial electric fields and plasma rotation are
tightly connected: it is considered that sheared $E\times B$ flows are
likely to reduce the edge turbulence level thus facilitating access to
High confinement (H) mode, see
Refs.~\cite{burrell1997shearflow,wagner2007hmode} and references
therein. A review on internal transport barriers and H mode in helical
systems can be found in Ref.~\cite{wagner2006barriers}. Both effects
have been measured at the heliac TJ-II~\cite{alejaldre1999first}:
transitions to core electron root confinement have been observed in
Electron Cyclotron Heated (ECH)
plasmas~\cite{castejon2002cerc}. Transitions to H mode have been
documented~\cite{sanchez2009transitions,estrada2009transitions}
together with mean and low frequency oscillating sheared $E\times B$
flows in plasmas heated by Neutral Beam Injection (NBI). \\

Neoclassical transport theory allows to predict the radial electric
field in helical devices by means of the ambipolarity
condition. Examples of these calculations exist for devices such as
W7-AS~\cite{kick1999w7as} and LHD~\cite{yokoyama2002lhd} among many
others~\cite{yokoyama2007cerc,maassberg1993stellarators}. Previous
neoclassical transport calculations of the ambipolar electric field at
TJ-II include ECH
plasmas~\cite{tribaldos2001nctj2,chmyga2002hibp,turkin2011predictive}
and also medium-density NBI plasmas~\cite{zurro2006rotation}. In
Refs.~\cite{chmyga2002hibp,zurro2006rotation}, the calculations were
compared with Heavy Ion Beam Probe (HIBP) and passive emission
spectroscopy measurements: qualitative agreement was obtained. A
number of additional HIBP measurements exist for ECH
plasmas~\cite{krupnik2005hibp,melnikov2005hibp,milligen2011transitions},
and more recently for NBI plasmas~\cite{melnikov2007hibp}. Near the
edge, the electric field has been measured by means of
reflectometry~\cite{estrada2006refle,estrada2009transitions} and, very
recently, by studying mode rotation
velocities~\cite{milligen2011islands}.

Ref.~\cite{tribaldos2001nctj2} includes a comprehensive study of the
transport coefficients and the flux balances for two ECH plasmas
(although no multiple roots were found, see below), but an analogous
work is missing for medium-density NBI plasmas, where only the
ambipolar electric field has been shown. Furthermore, lithium wall
coating of TJ-II has recently allowed~\cite{sanchez2009transitions}
transitions to regimes of relatively high density in NBI plasmas, and
these plasmas have not yet been described from the neoclassical
transport point of view. Additionally, many of the effects reported
above show a dependence on the magnetic
configuration~\cite{estrada2004cerc,estrada2009transitions}. No
qualitative changes are expected in the neoclassical radial transport
of these configurations since the main Fourier components of the
magnetic field strength do not change too
much~\cite{solano1988tj-ii,tribaldos2001nctj2}. Still, the flexibility
of TJ-II allows for exploring a large set of configurations (on a
shot-to-shot basis~\cite{alejaldre1999confinement} or
continuously~\cite{lopez2009cmode}) and a general study of the
variation of the transport coefficients and the neoclassical balance
may be of interest. More precise calculations for selected discharges
are underway. \\

%In this work, we aim to complete the previous neoclassical transport
%studies in view of the recent upgrades in TJ-II operation. The basic
%theory is reviewed in Section~\ref{SEC_CALCULATION}. The 100\_44\_64
%magnetic configuration, the most usually operated at TJ-II, is briefly
%described in Section~\ref{SEC_MONO}. Then its monoenergetic radial
%transport coefficient is discussed. Convolution and solution of the
%ambipolar equation yield the radial fluxes and the radial electric
%field for three plasmas: low density (ECH heated) and medium and high
%density (NBI heated). A Monte-Carlo technique for error propagation
%allows us to account for the convergence problems of
%DKES~\cite{hirshman1986dkes} for the long-mean-free-path ({\em lmfp}),
%enhanced by the complexity of the magnetic configuration of
%TJ-II. This is done in Section~\ref{SEC_BALANCE}. Finally we have
%explored part of the set of magnetic configurations of TJ-II with DKES
%calculations. In Section~\ref{SEC_CONFS} we compare the main Fourier
%coefficients describing the equilibria, show how the monoenergetic
%radial transport coefficients depends on the configuration and finally
%discuss the implications on the radial particle
%balance. Section~\ref{SEC_CONCLUSIONS} is devoted to the conclusions.
%\\
In this work, we aim to complete the previous neoclassical transport
studies in view of the recent upgrades in TJ-II operation: we study a
low-density ECH plasma and discuss, for the first time in detail from
the neoclassical transport point of view, the issue of multiple roots
at TJ-II. We also study a high-density NBI plasma and show for the
first time the radial fluxes. Finally, we extend the calculations in
these two plasmas to seven other configurations usually operated at
TJ-II. The paper is organized as follows: the basic theory is reviewed
in Section~\ref{SEC_CALCULATION}. The monoenergetic radial transport
coefficient for the 100\_44\_64 magnetic configuration, the most
usually operated at TJ-II, is briefly described in
Section~\ref{SEC_MONO}. Then, convolution and solution of the
ambipolar equation yield the radial fluxes and the radial electric
field for the two plasmas. A Monte-Carlo technique for error
propagation allows us to account for the convergence problems of
DKES~\cite{hirshman1986dkes} for the long-mean-free-path ({\em lmfp}),
enhanced by the complexity of the magnetic configuration of
TJ-II. This is done in Section~\ref{SEC_BALANCE}. Finally we have
explored part of the set of magnetic configurations of TJ-II with DKES
calculations. In Section~\ref{SEC_CONFS} we compare the main Fourier
coefficients describing the equilibria, show how the monoenergetic
radial transport coefficient depends on the configuration and finally
discuss the implications on the radial particle
balance. Section~\ref{SEC_CONCLUSIONS} is devoted to the conclusions.
\\

\section{Determination of the neoclassical radial fluxes}\label{SEC_CALCULATION}

The radial electric field $E_\mathrm{r}$ may be obtained from the
ambipolar condition of the neoclassical radial particle fluxes, which
in a pure plasma composed of electrons and ions reads:
\begin{equation}
\Gamma_\mathrm{e}(E_\mathrm{r})=\Gamma_\mathrm{i}(E_\mathrm{r})\,.\label{EQ_AMB}\\
\end{equation}
The neoclassical fluxes are linear combinations of the density and
temperature gradients and the radial electric
field~\cite{helander2002collisional,wakatani1998stellarator}. For each
species $\mathrm{b}$, the flux-surface-averaged radial particle flux
$\Gamma_\mathrm{b}$ and radial energy flux $Q_\mathrm{b}$ are:
\begin{eqnarray}
 \frac{\Gamma_\mathrm{b}}{n}&=&-
 L_{11}^\mathrm{b}\left(\frac{1}{n}\frac{\mathrm{d}n}{\mathrm{d} r} -
 Z_\mathrm{b}
 e\frac{E_\mathrm{r}}{T_\mathrm{b}}-\frac{3}{2}\frac{1}{T_\mathrm{b}}\frac{\mathrm{d}
 T_\mathrm{b}}{\mathrm{d} r}\right) -
 L_{12}^b\frac{1}{T_\mathrm{b}}\frac{\mathrm{d}
 T_\mathrm{b}}{\mathrm{d} r}\,,\label{EQ_LINEAR1}\\
 \frac{Q_\mathrm{b}}{n T_\mathrm{b}} &=& -
 L_{21}^\mathrm{b}\left(\frac{1}{n}\frac{\mathrm{d}n}{\mathrm{d} r} -
 Z_\mathrm{b}
 e\frac{E_\mathrm{r}}{T_\mathrm{b}}-\frac{3}{2}\frac{1}{T_\mathrm{b}}\frac{\mathrm{d}
 T_\mathrm{b}}{\mathrm{d} r}\right) -
 L_{22}^b\frac{1}{T_\mathrm{b}}\frac{\mathrm{d}
 T_\mathrm{b}}{\mathrm{d} r}\,.\label{EQ_LINEAR2}
\label{EQ_LINEAR}
\end{eqnarray}
Here $n$ is the density, and $T_\mathrm{b}$ and $Z_\mathrm{b}e$ are
the temperature and the charge. We consider current free operation,
$E_\parallel\!=\!0$. The thermal coefficients
$L^\mathrm{b}_{\mathrm{j}\mathrm{k}}$ at each radial position can be
calculated by convolution with a Maxwellian distribution of the
monoenergetic radial diffusion coefficient:
\begin{equation}
 L^\mathrm{b}_{\mathrm{j}\mathrm{k}}(r,n,T_\mathrm{i},T_\mathrm{e},E_\mathrm{r})\!=\!\frac{2}{\sqrt{\pi}}\int_0^\infty\mathrm{d}\,x^2\,e^{-x^2}x^{1+2(\delta_{\mathrm{j},2}+\delta_{\mathrm{k},2})}
 D_{11}(r,\nu^*,\Omega)\,.
\label{EQ_CONVOLUTION}
\end{equation}
The integration variable,
$x\!=\!v/v_{\mathrm{t}\mathrm{h}}^\mathrm{b}$, is the particle
velocity normalized by the thermal velocity. The monoenergetic radial
transport coefficient $D_{11}$, which we discuss in
Section~\ref{SEC_MONO}, depends on the collisionality $\nu^*\!=\!\nu
R/v\iota$ and the {\em electric field parameter}
$\Omega\!\equiv\!E_\mathrm{r}/(vB_0)$. Here, $\nu$ is the collision
frequency, $\iota$ is the rotational transform, $R$ the major radius
and $B_0$ is the (0,0) Fourier harmonic of the magnetic field strength
in Boozer coordinates.\\

From Eqs.~(\ref{EQ_AMB}),~(\ref{EQ_LINEAR1}),~(\ref{EQ_LINEAR2})
and~(\ref{EQ_CONVOLUTION}), it is clear that the determination of the
radial electric field in helical plasmas is a non-linear problem due
to the $L_{\mathrm{j}\mathrm{k}} (E_\mathrm{r})$ dependence. This may
lead to several solutions or {\em roots} of the ambipolar
condition~\cite{mynick1983roots} when solving the ambipolar equation
by means of root-finding algorithms. When this happens, we select
among ion root $E_\mathrm{r}^\mathrm{i}$ and electron root
$E_\mathrm{r}^\mathrm{e}$ according to a thermodynamic condition:
minimization of the generalized heat production rate due to
neoclassical transport, see
e.g.~\cite{turkin2011predictive,maassberg1993root}. $I\!<\!0$
($I\!>\!0$) means that the electron (ion) root is selected, where
\begin{equation}
I=\int_{E_\mathrm{r}^\mathrm{i}}^{E_\mathrm{r}^\mathrm{e}}(\Gamma_\mathrm{i}-\Gamma_\mathrm{e})\mathrm{d}E_\mathrm{r}\,.
\end{equation}
Note that the transition region between roots has been imposed to have
zero radial width. Another option would have been to solve a diffusion
equation for $E_\mathrm{r}$, instead of Eq.~(\ref{EQ_AMB}).

For error estimate, we follow the Monte-Carlo method described in
Ref.~\cite{velasco2011bootstrap}: we start with a database of the
radial diffusion coefficient $D_{11}(r,\nu^*,\Omega)$ with the
corresponding error bars. For every value of $r$, $\nu^*$ and
$\Omega$, we give a numerical value to $D_{11}$ in
Eq.~(\ref{EQ_CONVOLUTION}) by employing a Gaussian random number and
then solve Eq.~(\ref{EQ_AMB}). By repeating this procedure a number of times, we obtain
averages and standard deviations of $E_\mathrm{r}$ and the other
relevant quantities. \\

It must be noted that the neoclassical ordering may be violated at
TJ-II~\cite{tribaldos2005global,velasco2009finite} if the widths of
the ion drift-orbits are large. This makes the diffusive picture fail,
and a convective term corresponding to ripple-trapped particles should
be added to $\Gamma_\mathrm{i}$ in Eq.~(\ref{EQ_AMB}), leading to a
smaller electric field. The higher the collisionality, the lower the
correction. The incompressibility of the $E\times B$ drift, one of the
approximations in the calculation~\cite{beidler2007icnts}, is valid
for the NBI plasmas, where the electric field is small in absolute
value, but it may fail when the electron root is realized and the
electric field is close to resonance values. This may also lead to
underestimate the ion flux, and thus to overestimate the positive
radial electric field. Finally, although the monoenergetic
calculations do not conserve momentum, momentum-correction is
negligible for the radial transport of non-quasisymmetric
stellarators~\cite{maassberg2009momentum,lore2010quasi}.

\section{Calculation and results}\label{SEC_RESULTS}

%\subsection{Magnetic configuration}\label{SEC_CONF}

%\begin{figure}
%\begin{center}
%\subfigure{\includegraphics[angle=270,scale=0.25]{blevel.ps}}
%\subfigure{\includegraphics[angle=270,scale=0.25]{bline.ps}}
%\end{center}
%\caption{Contour lines of constant $B$ at $\rho\!=\!0.7$ (left) ; red
%corresponds to $B/B_0\!>\!1$ and green to $B/B_0\!<\!1$; consecutive
%lines differ by $B/B_0\!=\!0.015$. $B$ along the field line starting at
%$\rho\!=\!0.7\,$, $\theta\!=\!0\,$ and $\phi\!=\!0$ (right).}
%\label{FIG_BLEVEL}
%\end{figure}

%In Fig.~\ref{FIG_BLEVEL}, we plot the contour lines of constant
%magnetic field strength $B$ at the flux-surface $\rho\!=\!0.7$. Here,
%$\rho\!=\!\sqrt{\psi/\psi_0}$ is the normalized radial coordinate,
%where $\psi$ and $\psi_0$ are the toroidal flux through the local and
%the last closed magnetic surfaces respectively. Within the same
%surface, the value of $B$ along the field line is plotted in
%Fig.~\ref{FIG_BLEVEL}, starting at the poloidal and toroidal angles
%$\theta\!=\!0\,$ and $\phi\!=\!0$. The broad spectrum of TJ-II appears
%here as a rich structure of maxima and minima. 

\subsection{The magnetic configuration and the radial diffusion monoenergetic coefficient}\label{SEC_MONO}

TJ-II is a four-period ($N\!=\!4$) flexible heliac of medium size,
with strong helical variation of the magnetic axis and magnetic
surfaces with bean-shaped cross-section and small Shafranov shift. The
so-called 100\_44\_64 configuration (the most often employed during
TJ-II operation) has a major radius of $R\!=\!1.504\,$m, its minor
radius is $a\!=\!0.192\,$m and its volume-averaged magnetic field is
$0.957\,$T. We stick to the vacuum equilibrium calculated using
VMEC~\cite{hirshman1986vmec}, which has a flat iota profile, with
$\iota(0)\!=\!-1.551$ and $\iota(a)\!=\!-1.650$.

The radial diffusion coefficient has been calculated with DKES, in an
independent simulation for each radial position and value of $\nu^*$
and $\Omega$. The distribution function has been described with up to
150 Legendre polynomials and 2548 Fourier modes. For the description
of each magnetic surface, the largest 50 Fourier modes have been
employed.

\begin{figure}
\begin{center}
\includegraphics[angle=270,width=1\columnwidth]{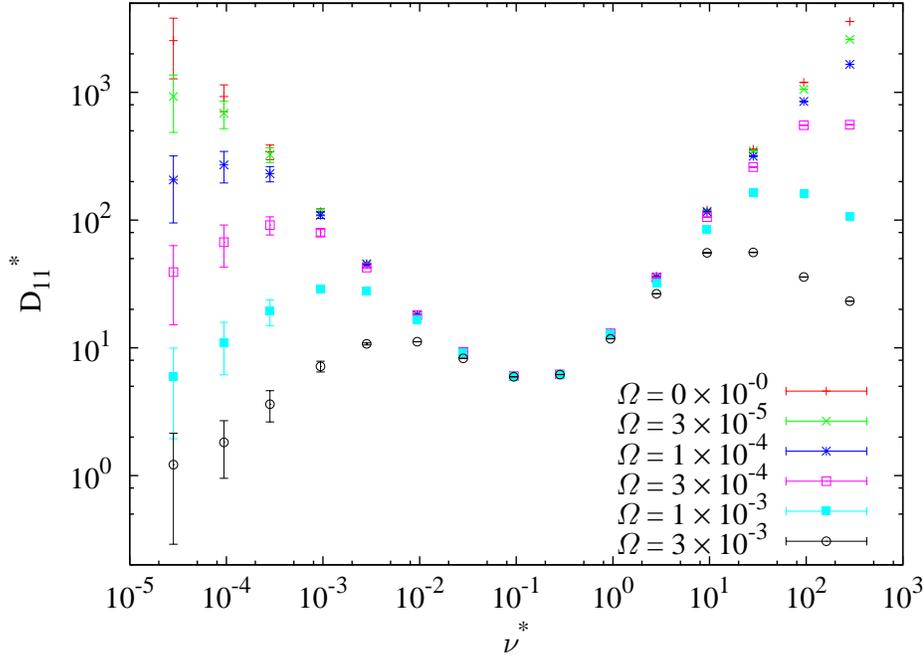}
\end{center}
\caption{Monoenergetic radial diffusion coefficient at $\rho\!=\!0.7$ as a
function of the collisionality for several values of the normalized electric field.}
\label{FIG_MONO}
\end{figure}

Fig.~\ref{FIG_MONO} shows the normalized radial diffusion coefficient
$D^*_{11}\!=\!D_{11}/D_{11}^\mathrm{p}$ calculated for several values
of the collisionality and the normalized electric field, at
$\rho\!=\!0.7$. The normalization $D_{11}^\mathrm{p}$ is the value of
$D_{11}$ in the plateau regime for the equivalent axisymmetric
tokamak, as in Ref.~\cite{beidler2011ICNTS}.

The Pfirsch-Schl\"uter (PS), plateau and long-mean-free-path ({\em
lmfp}) regimes~\cite{wakatani1998stellarator} are clearly visible in
Fig.~\ref{FIG_MONO}, with a qualitative dependence on $\nu^*$ and
$\Omega$ equal to that of the classical stellarator. The same
behaviour has been reported for similar configurations of TJ-II and
other stellarators~\cite{tribaldos2001nctj2,beidler2011ICNTS}. This
{\em text-book} dependence of $D^*_{11}$ with $\nu^*$ and $\Omega$ can
be summarized as follows:
\begin{itemize}
\item $\nu$ regime for high collisionalities, with reduction of
transport for large radial electric fields due to $E\times B$ poloidal
precession~\cite{igitkhanov2006impurity}.
\item Independence from $\nu^*$ and $\Omega$ for intermediate
collisionality.
\item $1/\nu$ dependence for low collisionalities and $\Omega\!=\!0$,
suppressed by the electric field, which leads to $\nu$ and
$\sqrt{\nu}$ regimes~\cite{galeev1979theory}.
\end{itemize}

The large error bars for collisionalities lower than $10^{-4}$ might
lead to inaccurate results for low-density plasmas, due to the $1/\nu$
dependence of the coefficient. In this work, these error bars have
been propagated to the final results following the method described in
Ref.~\cite{velasco2011bootstrap}: as we will see in
Section~\ref{SEC_BALANCE}, the collisionality is usually high enough
so that this procedure yields accurate results.

A database of $D^*_{11}$ has been built in the
($\rho$,$\nu^*$,$\Omega$)-space, with $\rho$ between 0.1 and 1,
$\nu^*$ between $3\times 10^{-5}$ and $3\times 10 ^2$ and $\Omega$
between 0 and $1\times 10^{-1}$. The convolution of
Eq.~(\ref{EQ_CONVOLUTION}) requires interpolation and extrapolation in
this three-dimensional database. The interpolation is done by means of
3-point Lagrange, with $\nu^*$, $\Omega$ and $D^*_{11}$ in logarithmic
scale. Since $D^*_{11}$ is calculated at several tens of radial
positions, interpolation in $\rho$ is not necessary (but note that
Eq.~(\ref{EQ_CONVOLUTION}) is local, so that radial interpolation
would not be required at this step). We have made use of the
asymptotic collisional and collisionless
limits~\cite{beidler2011ICNTS} where possible. Neither the choice of
interpolation algorithm nor the extrapolation procedure affect
significantly the final results~\cite{spong2005flow}. Integration has
been made by means of Gauss-Laguerre of order 64 (order 200 yields
compatible results).

\subsection{Radial balances}\label{SEC_BALANCE}

%We calculate the ambipolar electric field for three regimes of TJ-II:
%a low-density ECH plasma
%(Figs.~\ref{FIG_PROFILES_LD},~\ref{FIG_THERMALLD},~\ref{FIG_FLUXES_LD},~\ref{FIG_AMB2D}
%and~\ref{FIG_AMB3DECH}) a medium-density NBI plasma
%(Figs.~\ref{FIG_PROFILES_MD},~\ref{FIG_THERMALMD},~\ref{FIG_FLUXES_MD}
%and~\ref{FIG_AMB3DNBI}) and a high-density NBI plasma
%(Figs.~\ref{FIG_PROFILES_HD},~\ref{FIG_THERMALHD},~\ref{FIG_FLUXES_HD}
%and~\ref{FIG_AMB3DNBI}). Since the walls are coated with
%lithium~\cite{tabares2008lithium}, TJ-II plasmas have low impurity
%content, and the effective charge is taken to be
%$Z_\mathrm{eff}\!=\!1\,$ everywhere.

\begin{figure}
\begin{center}
\subfigure{\includegraphics[angle=270,scale=0.25]{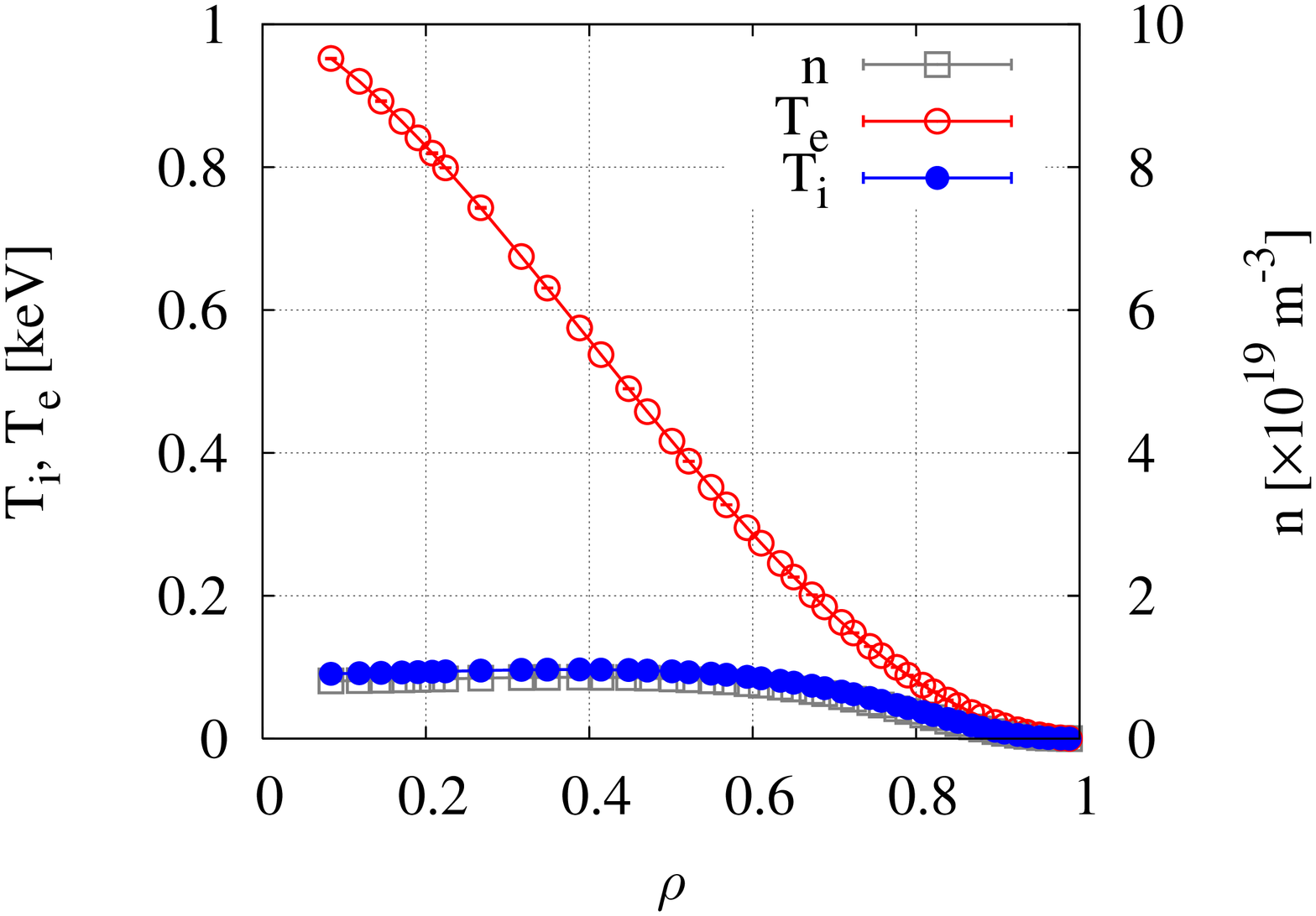}}
\subfigure{\includegraphics[angle=270,scale=0.25]{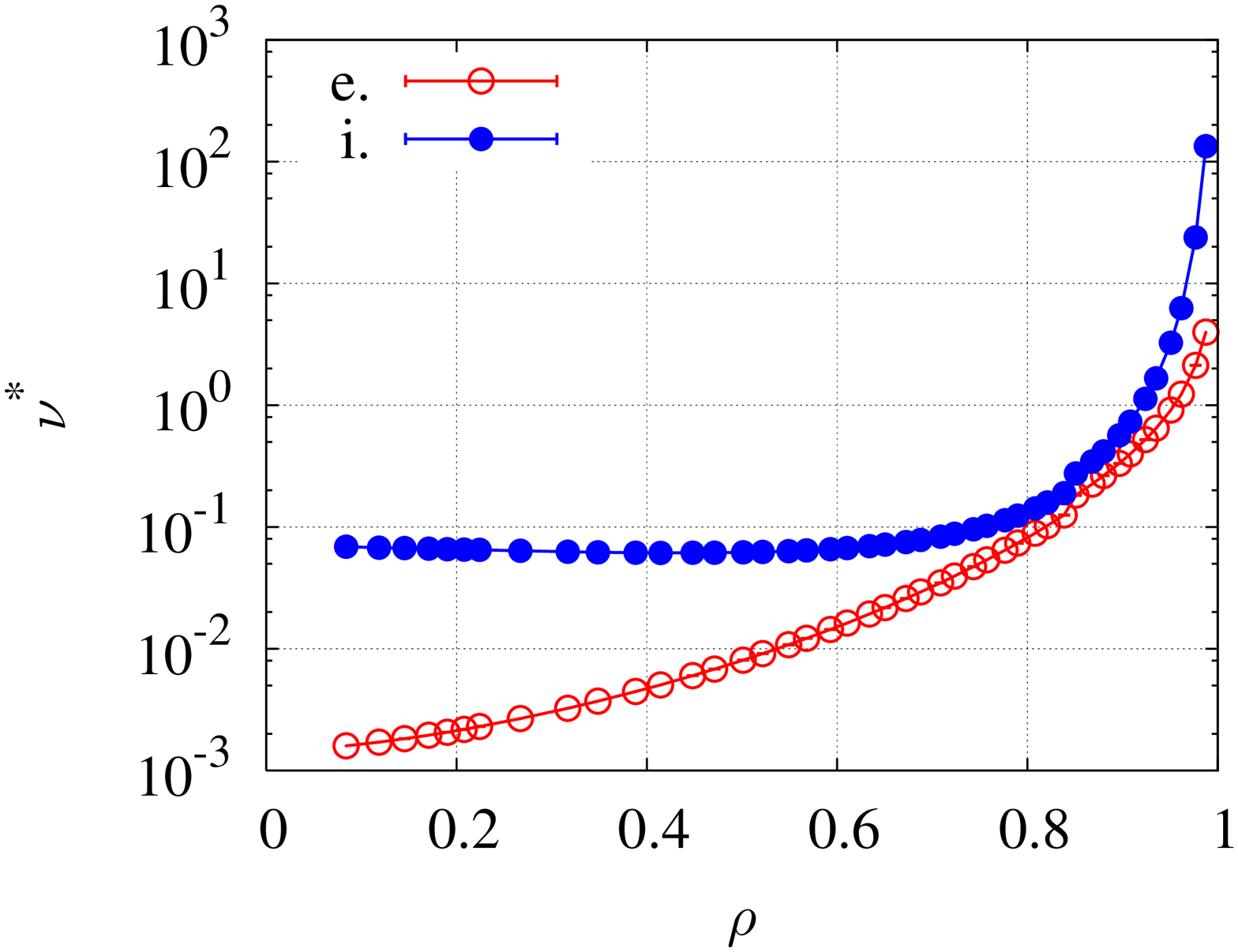}}
\end{center}
\caption{Plasma profiles for the low-density plasma: density and temperatures (left); collisionalities (right).}
\label{FIG_PROFILES_LD}
\end{figure}

\begin{figure}
\begin{center}
\subfigure{\includegraphics[angle=270,scale=0.25]{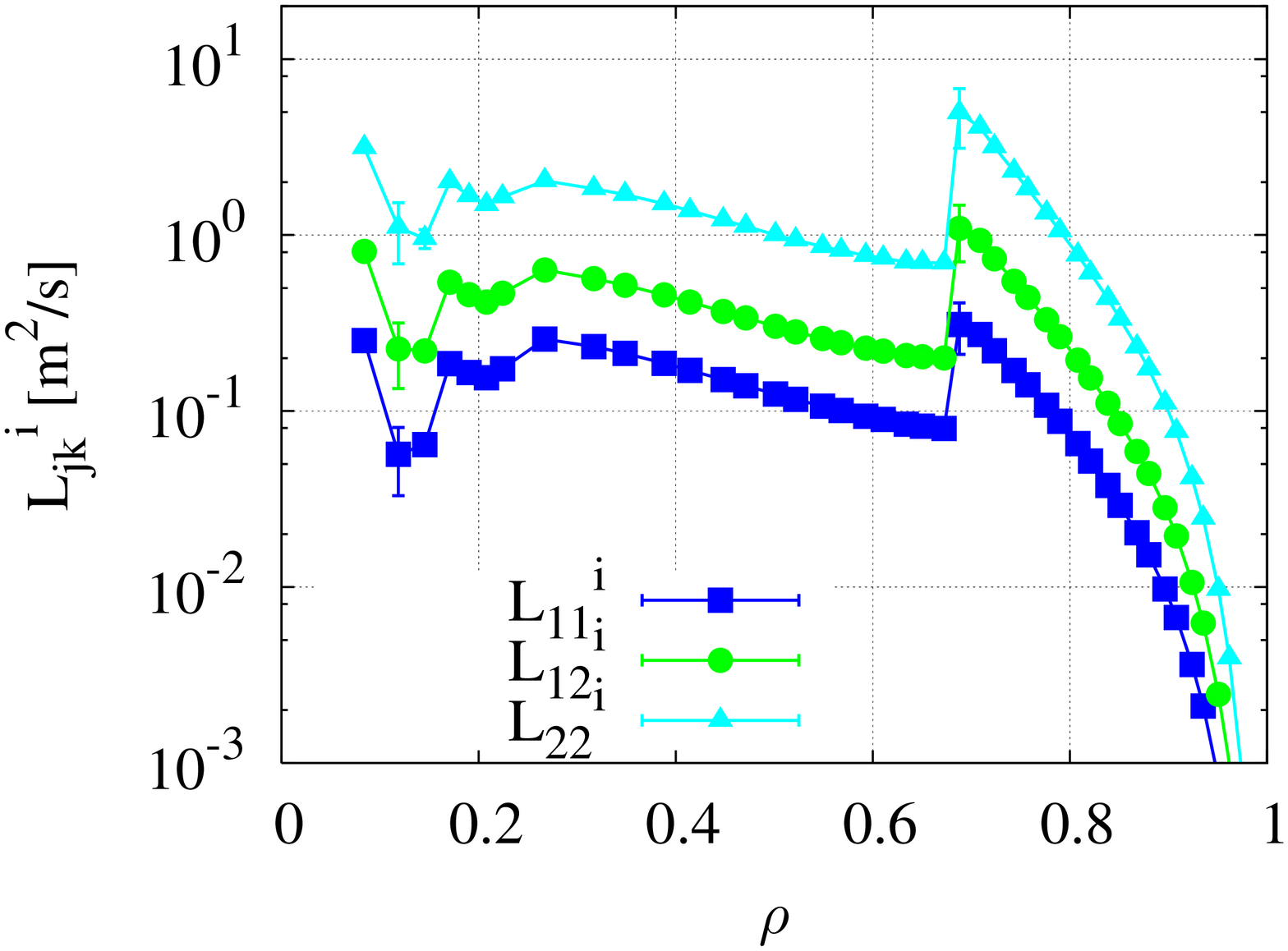}}
\subfigure{\includegraphics[angle=270,scale=0.25]{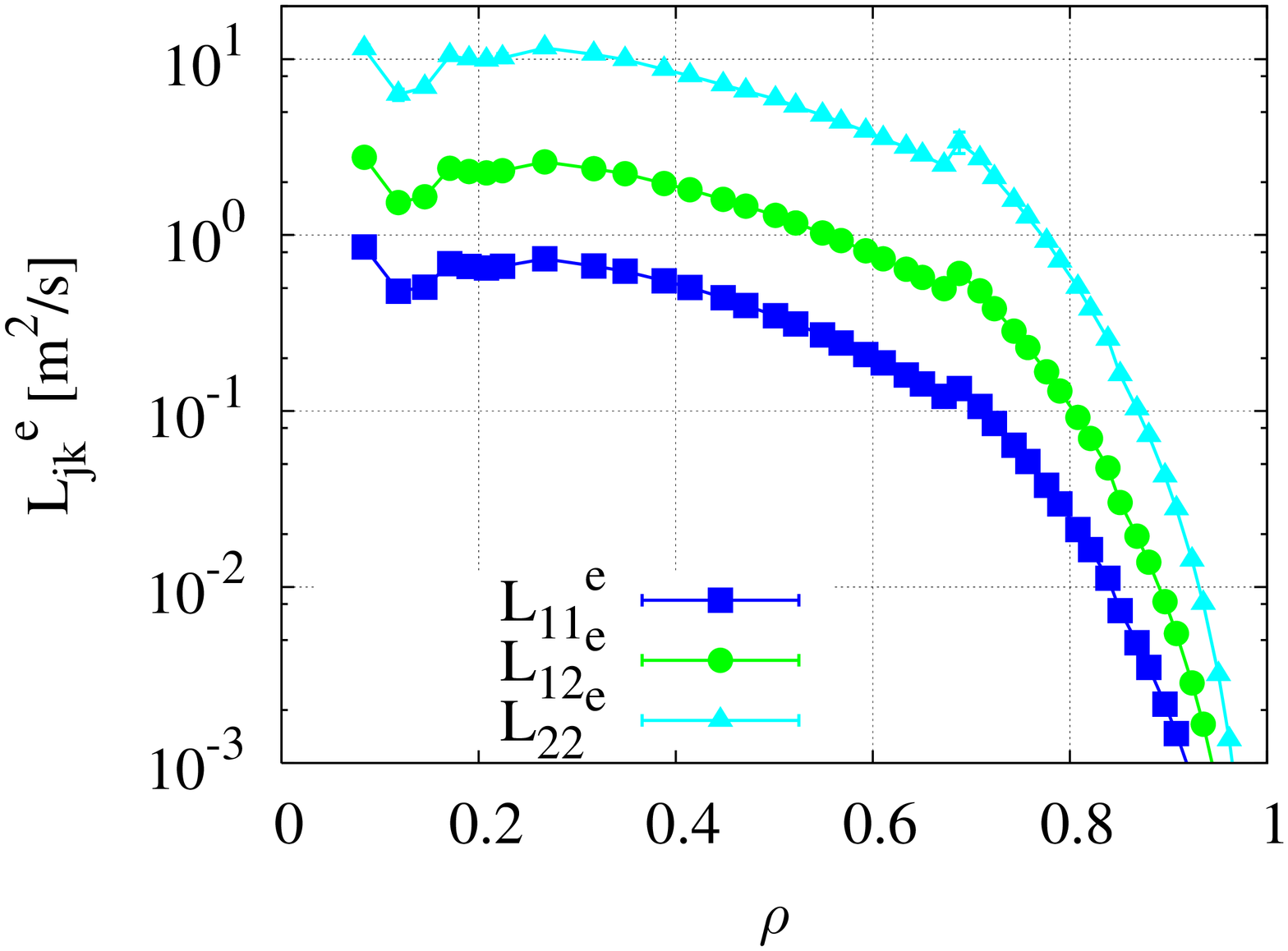}}
\end{center}
\caption{Thermal transport coefficients for the low-density plasma.}
\label{FIG_THERMALLD}
\end{figure}

\begin{figure}
\begin{center}
\subfigure{\includegraphics[angle=270,scale=0.25]{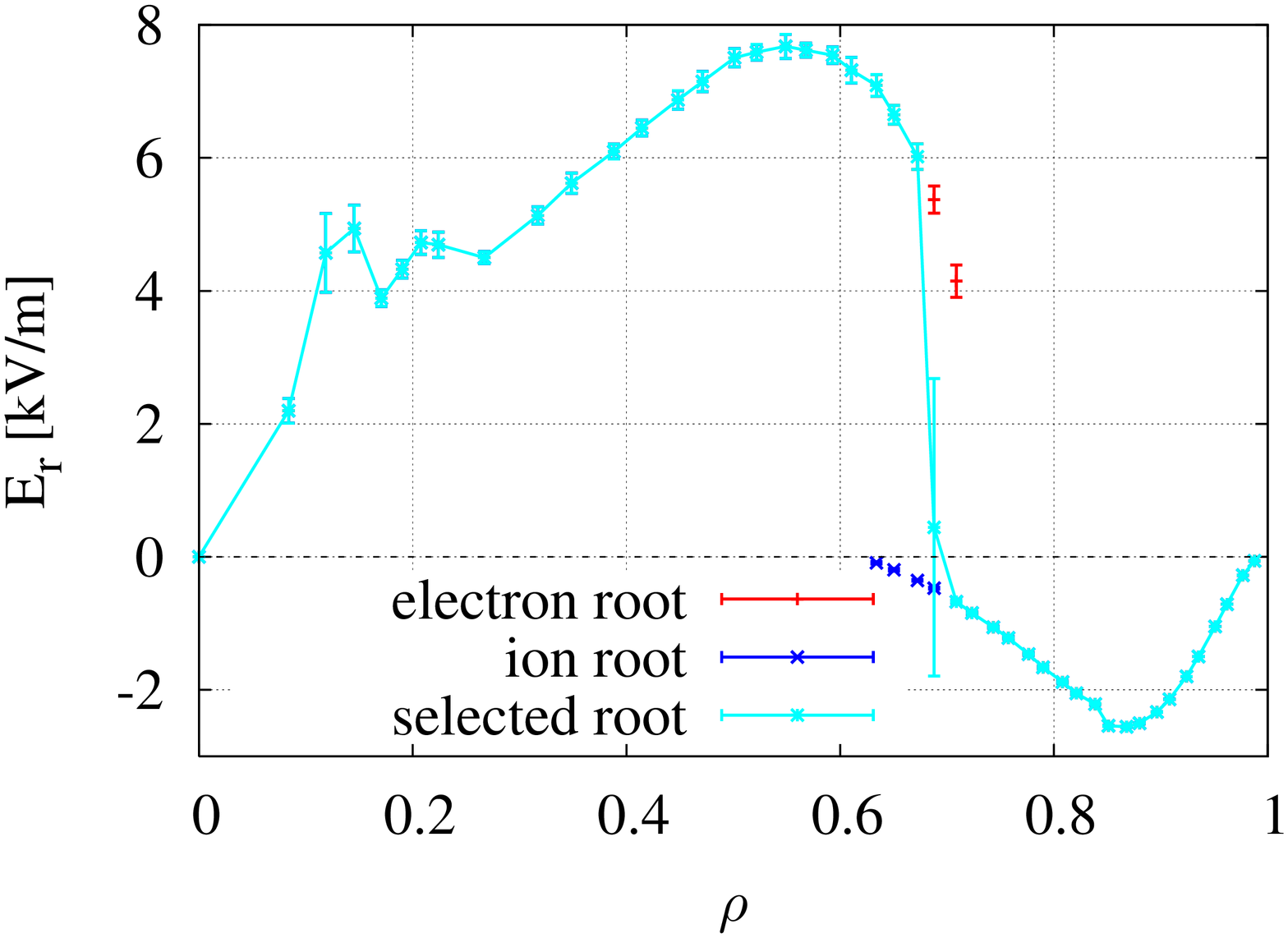}}
\subfigure{\includegraphics[angle=270,scale=0.25]{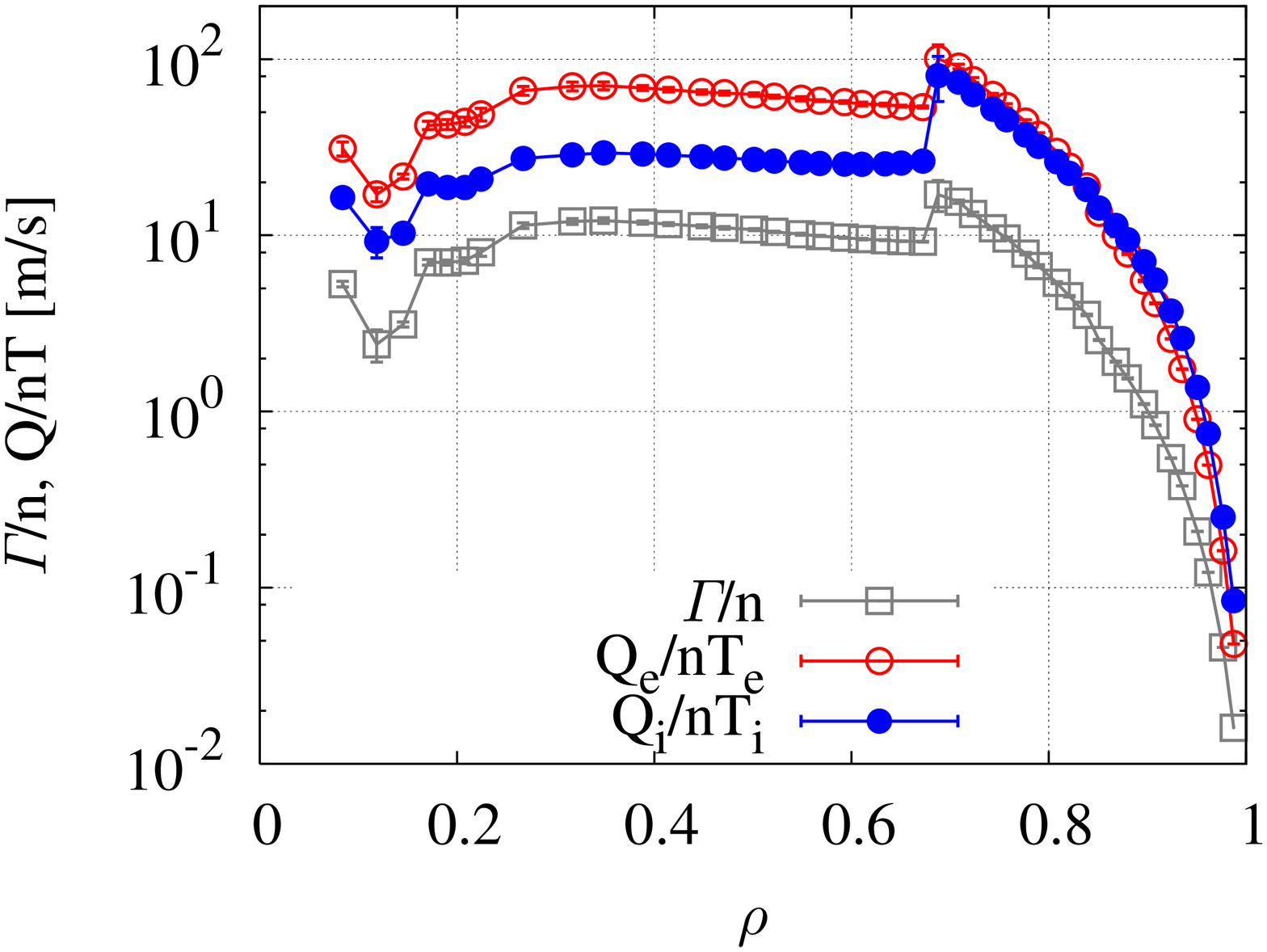}}
\end{center}
\caption{Radial balance for the low-density plasma: ambipolar radial electric field (left); ambipolar particle and energy fluxes (right).}
\label{FIG_FLUXES_LD}
\end{figure}

\begin{figure}
\begin{center}
\subfigure{\includegraphics[angle=270,scale=0.25]{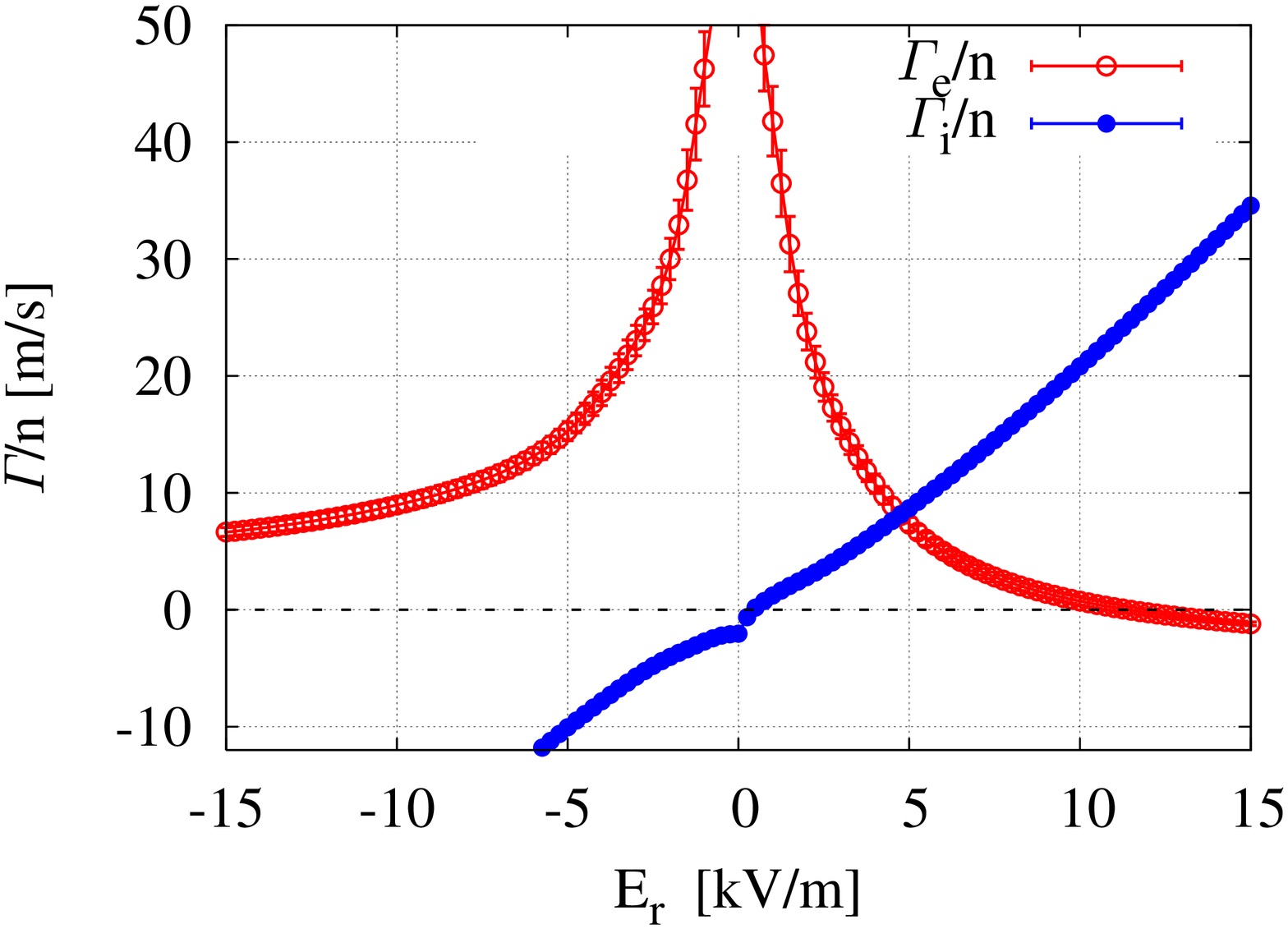}}
\subfigure{\includegraphics[angle=270,scale=0.25]{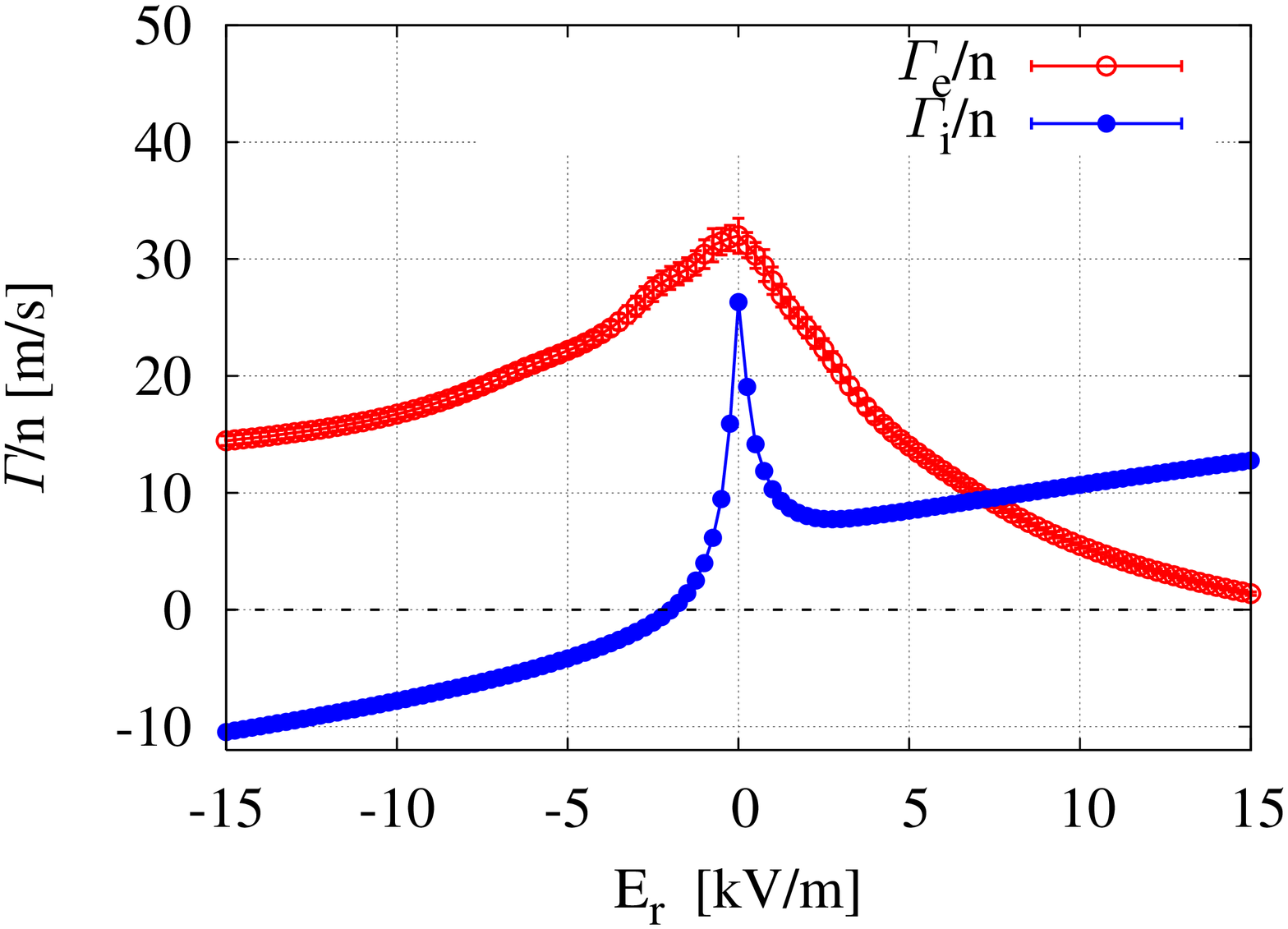}}
\subfigure{\includegraphics[angle=270,scale=0.25]{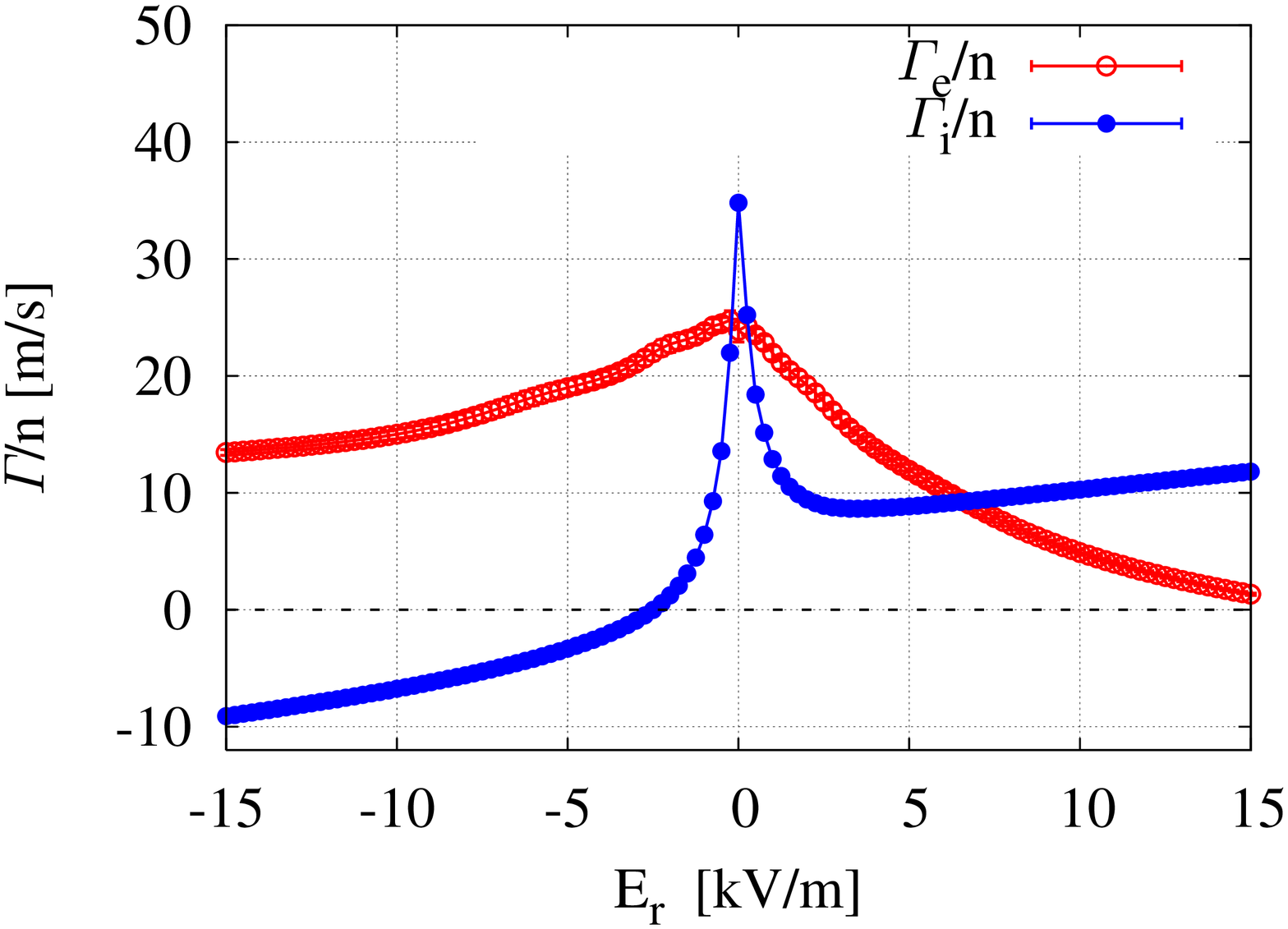}}
\subfigure{\includegraphics[angle=270,scale=0.25]{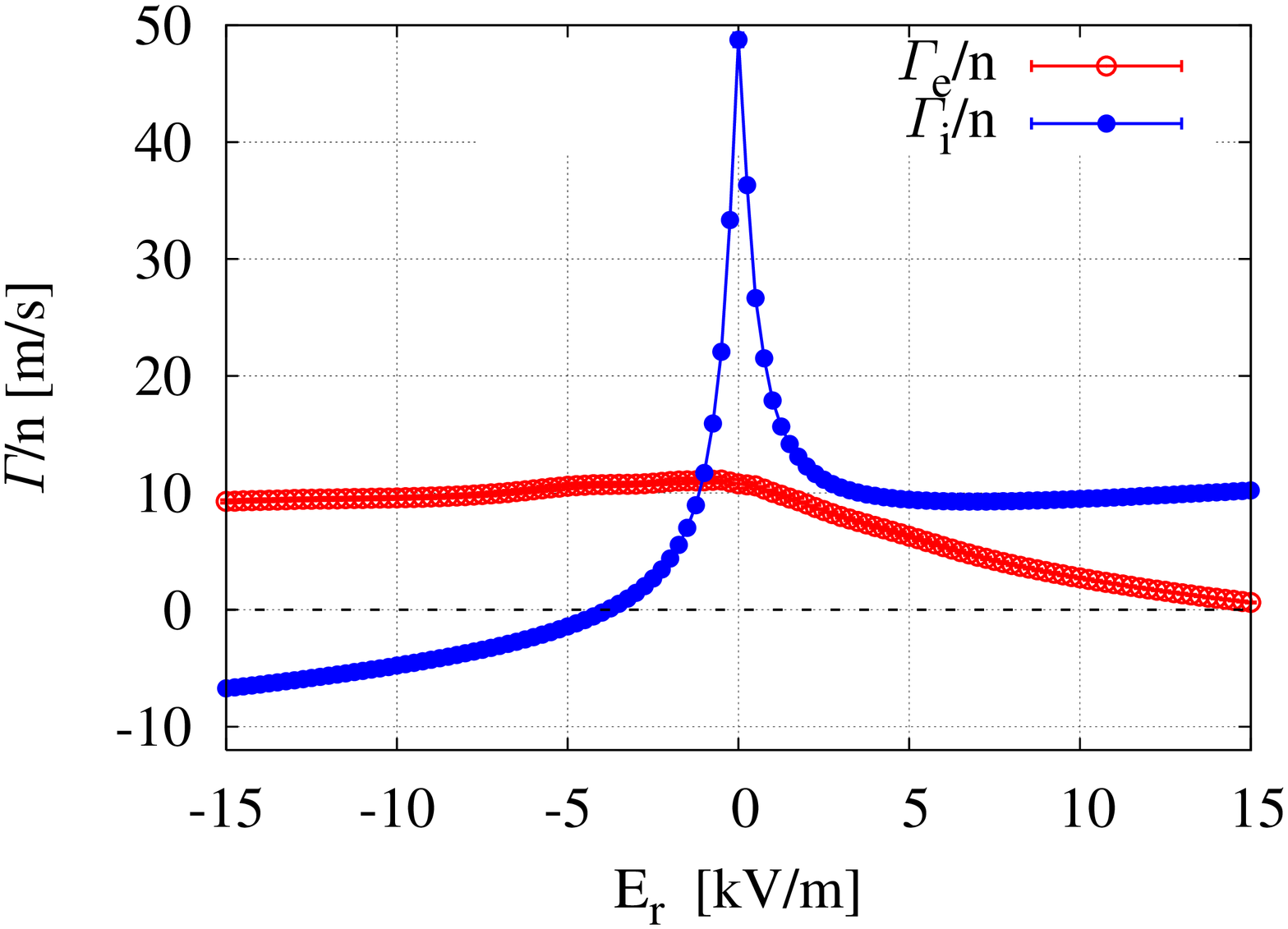}}
\end{center}
\caption{Radial particle fluxes as a function of the radial electric
field at 4 radial positions of the ECH plasma: $\rho\!=\!0.22$ (top
left), $\rho\!=\!0.61$ (top right), $\rho\!=\!0.65$ (bottom left) and
$\rho\!=\!0.74$ (bottom right).}
\label{FIG_AMB2D}
\end{figure}

\begin{figure}
\begin{center}
\includegraphics[angle=270,width=\columnwidth]{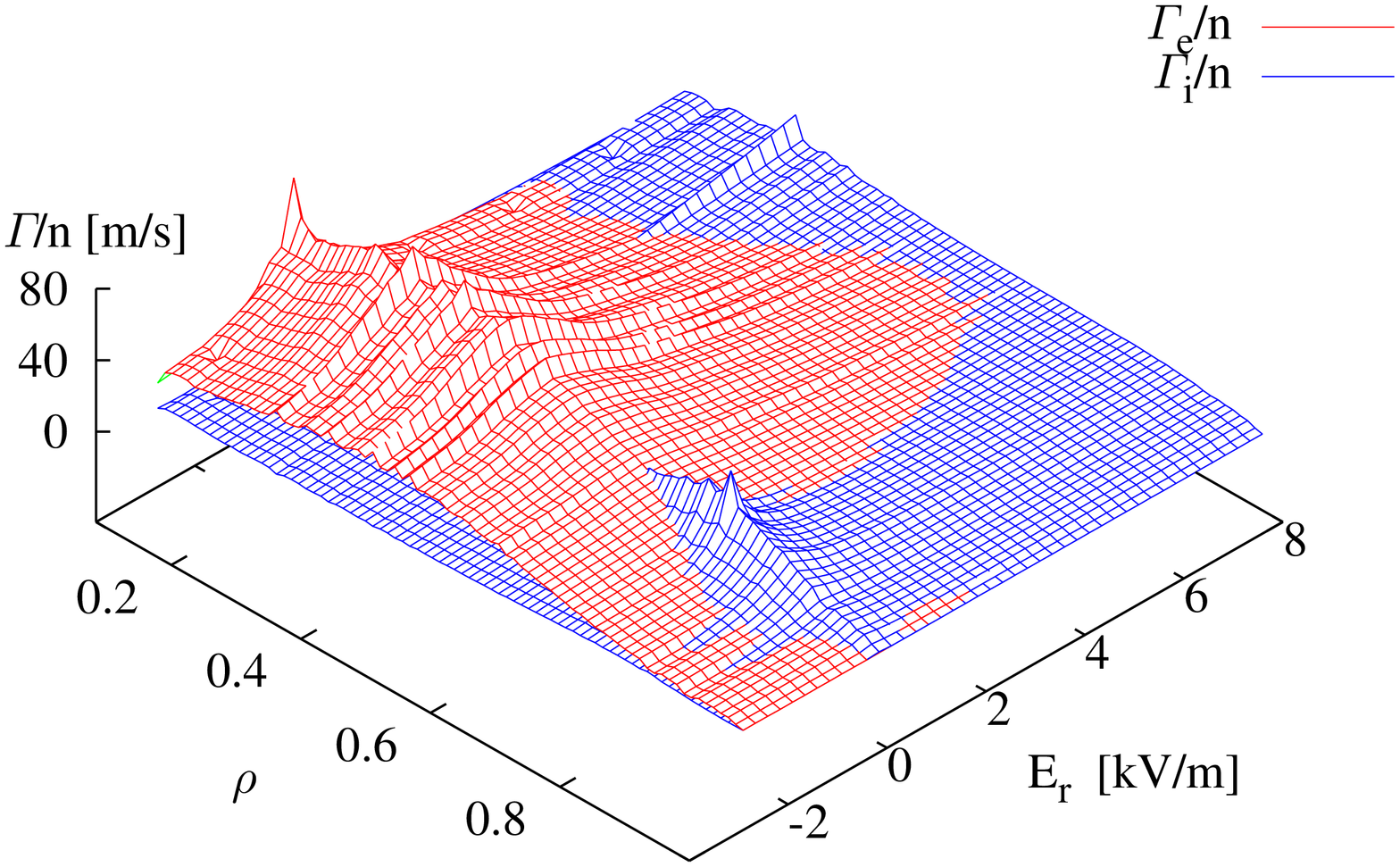}
\end{center}
\caption{Radial profile of the solution of the ambipolarity equation for the ECH plasma.}
\label{FIG_AMB3DECH}
\end{figure}

\begin{figure}
\begin{center}
\subfigure{\includegraphics[angle=270,scale=0.25]{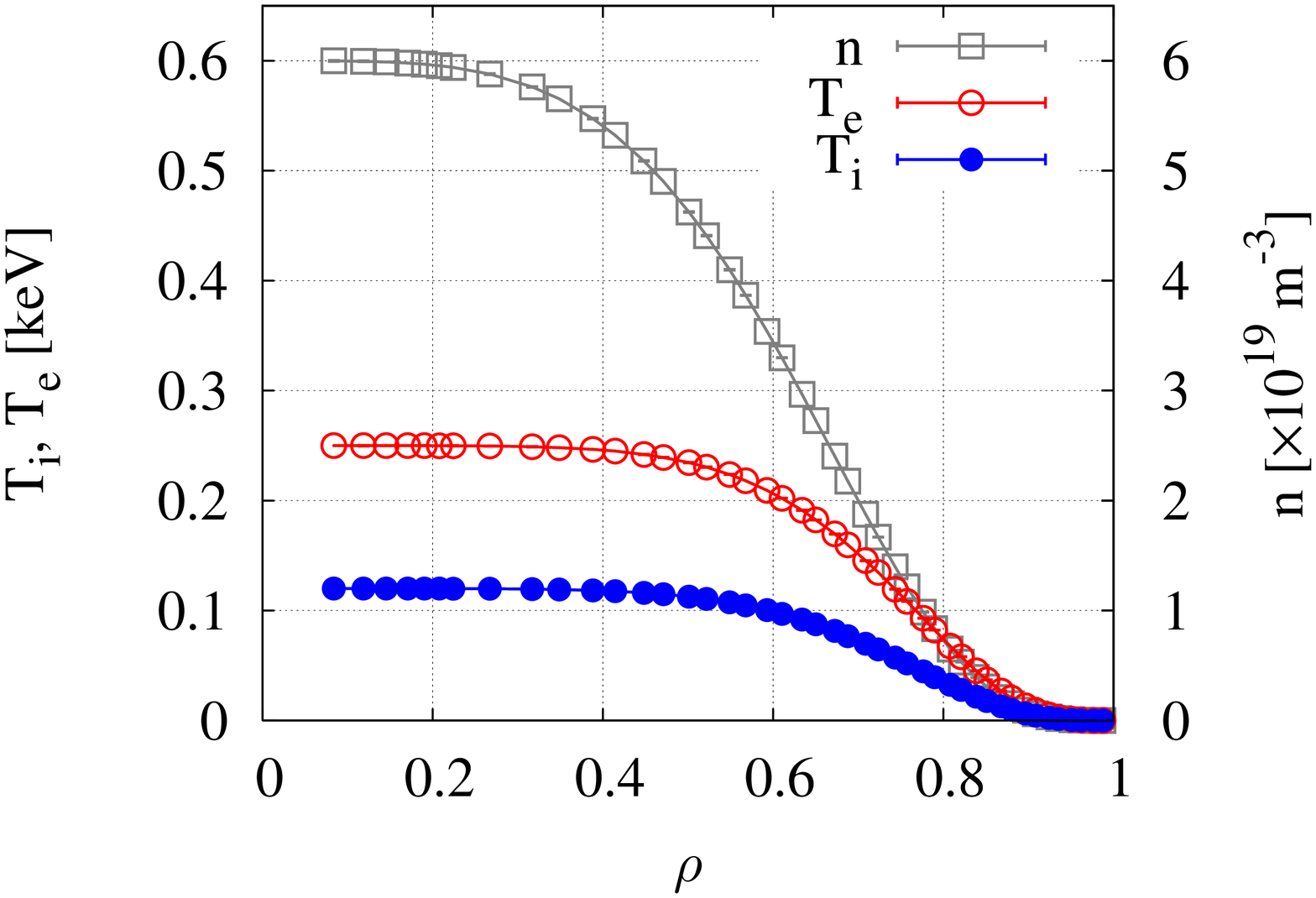}}
\subfigure{\includegraphics[angle=270,scale=0.25]{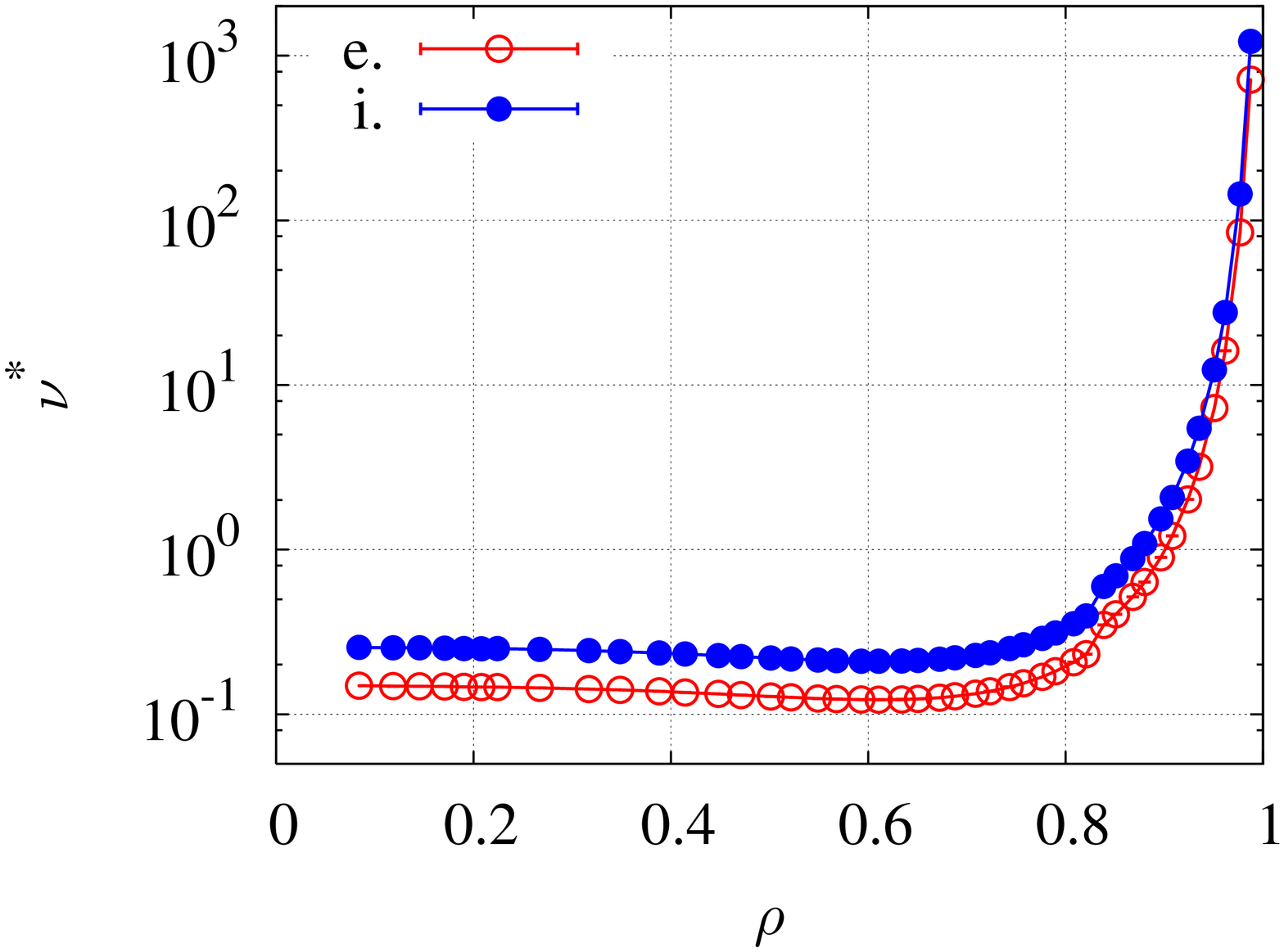}}
\end{center}
\caption{Plasma profiles for the high-density plasma: density and temperatures (left); collisionalities (right).}
\label{FIG_PROFILES_HD}
\end{figure}

\begin{figure}
\begin{center}
\subfigure{\includegraphics[angle=270,scale=0.25]{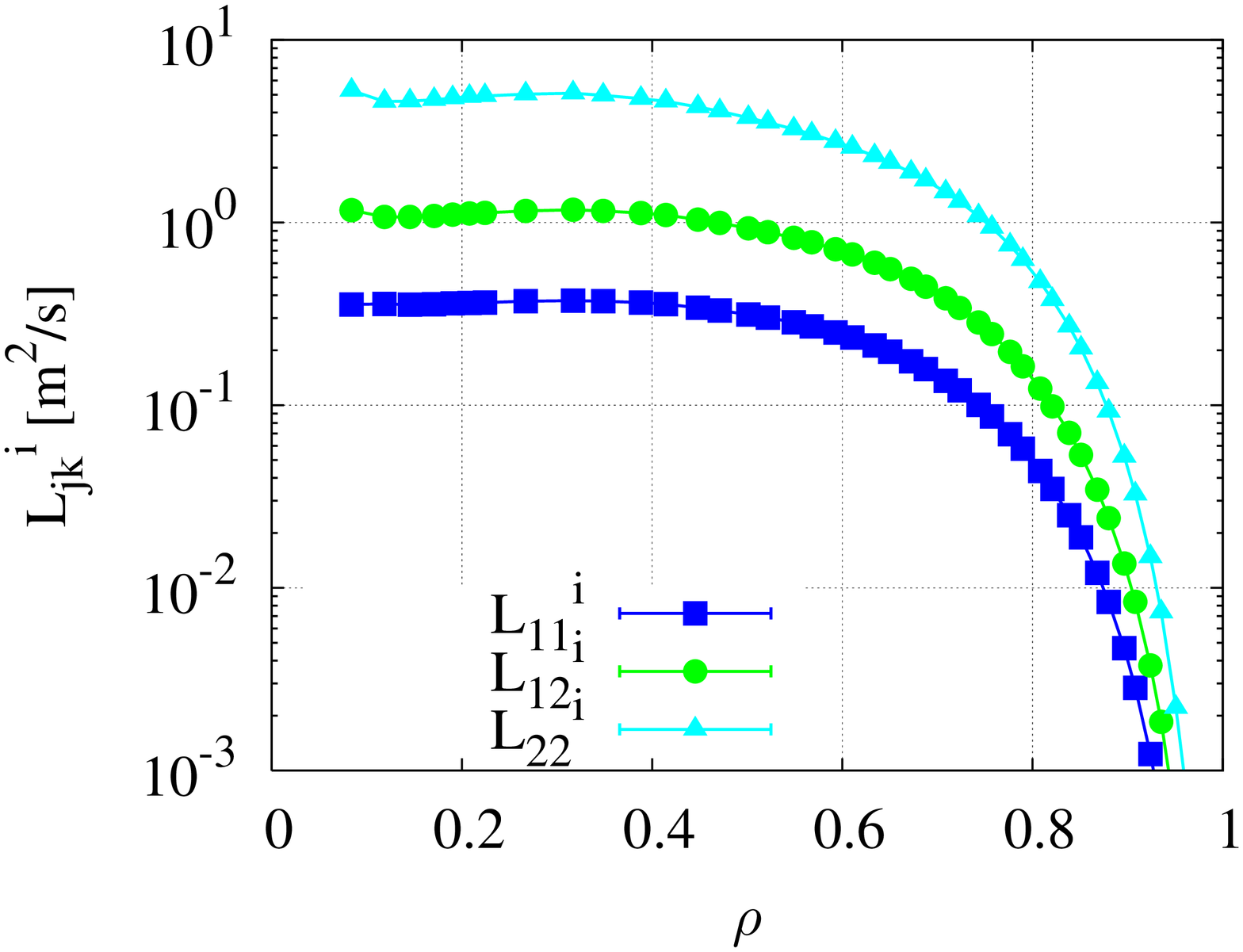}}
\subfigure{\includegraphics[angle=270,scale=0.25]{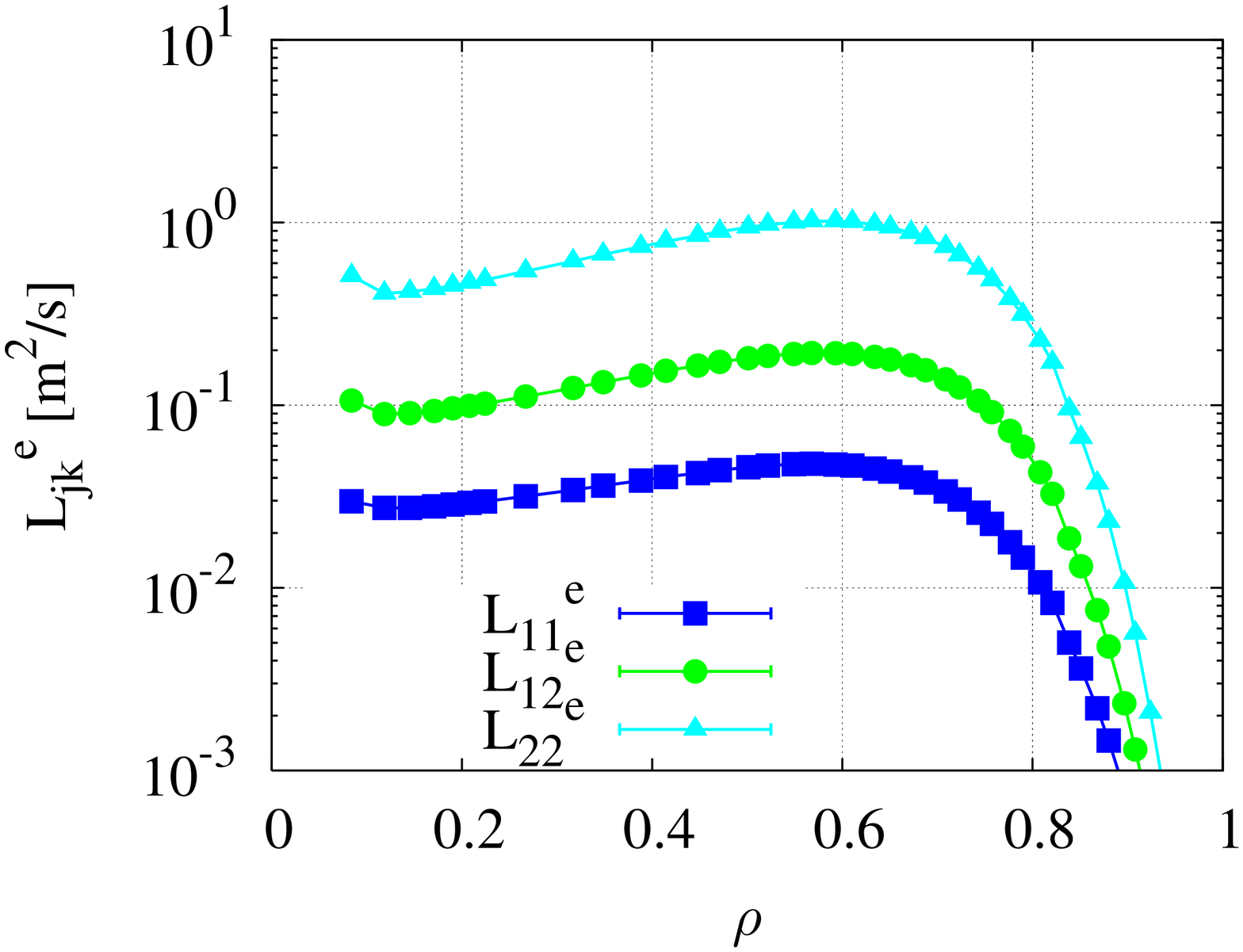}}
\end{center}
\caption{Thermal transport coefficients for the high-density plasma.}
\label{FIG_THERMALHD}
\end{figure}

\begin{figure}
\begin{center}
\subfigure{\includegraphics[angle=270,scale=0.25]{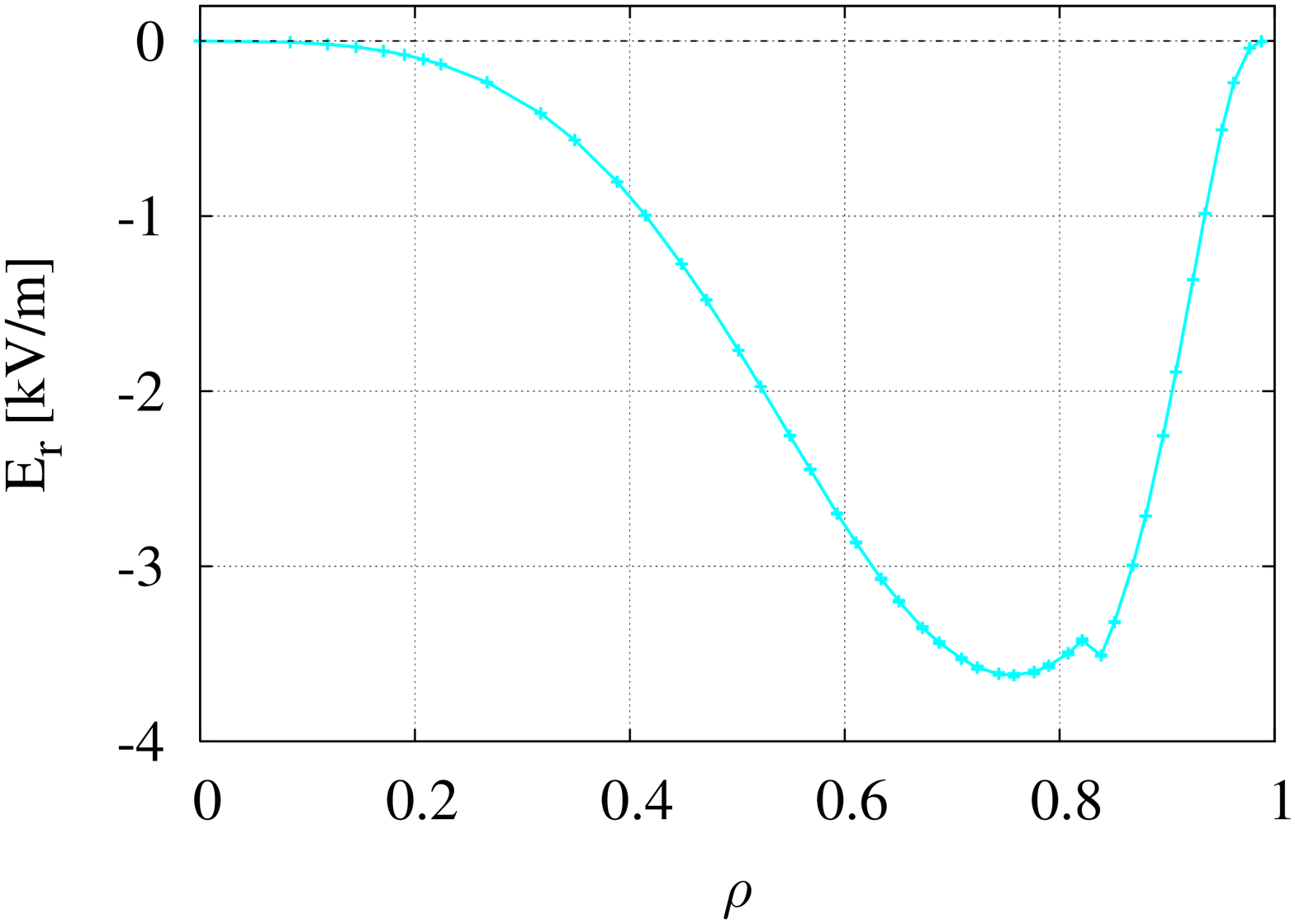}}
\subfigure{\includegraphics[angle=270,scale=0.25]{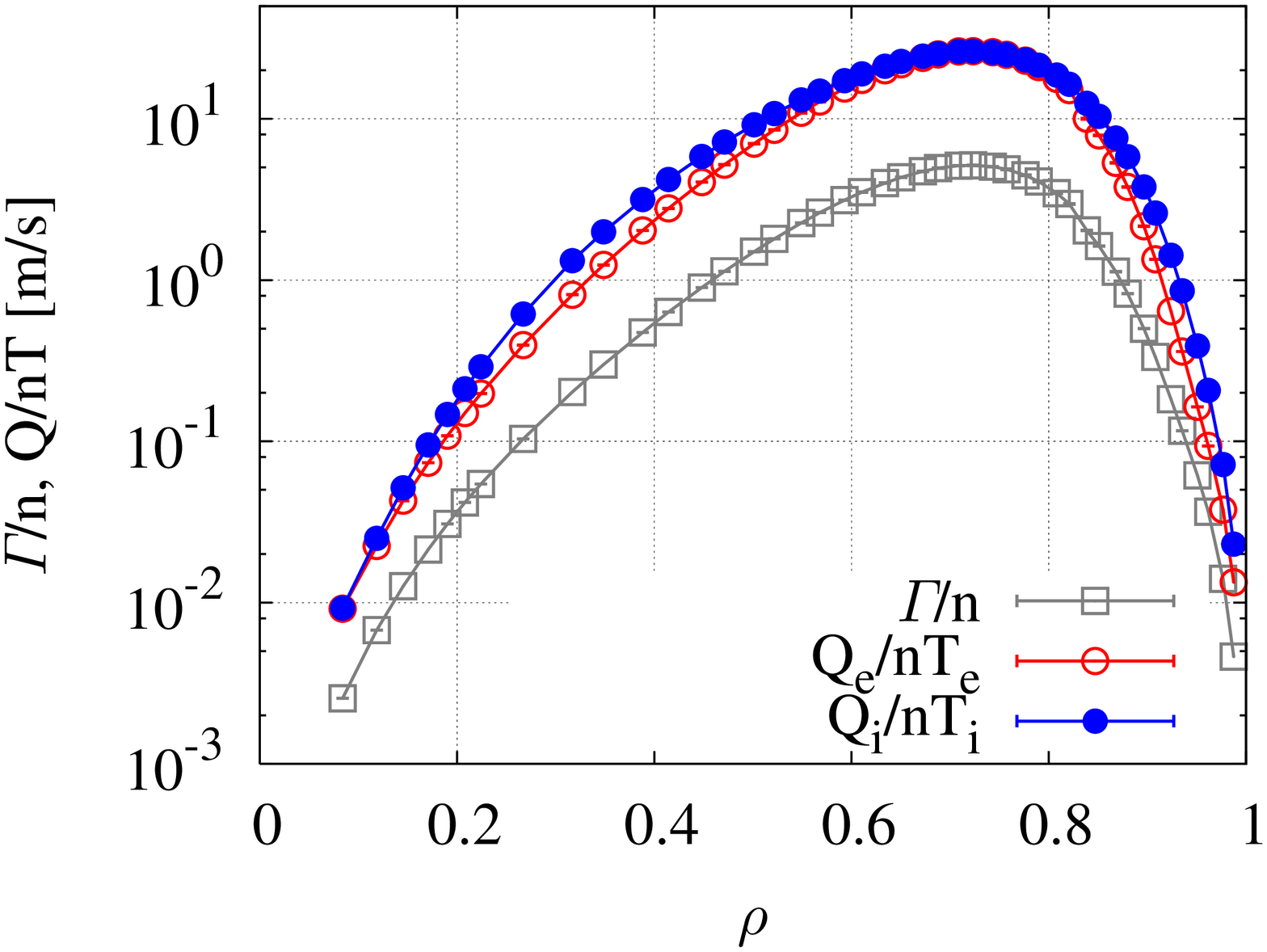}}
\end{center}
\caption{Radial balance for the high-density plasma: ambipolar radial electric field (left); ambipolar particle and energy fluxes (right).}
\label{FIG_FLUXES_HD}
\end{figure}

\begin{figure}
\begin{center}
%\subfigure{\includegraphics[angle=270,scale=0.25]{amb3dnbi1.ps}}
%\subfigure{\includegraphics[angle=270,scale=0.25]{amb3dnbi2.ps}}
\includegraphics[angle=270,width=\columnwidth]{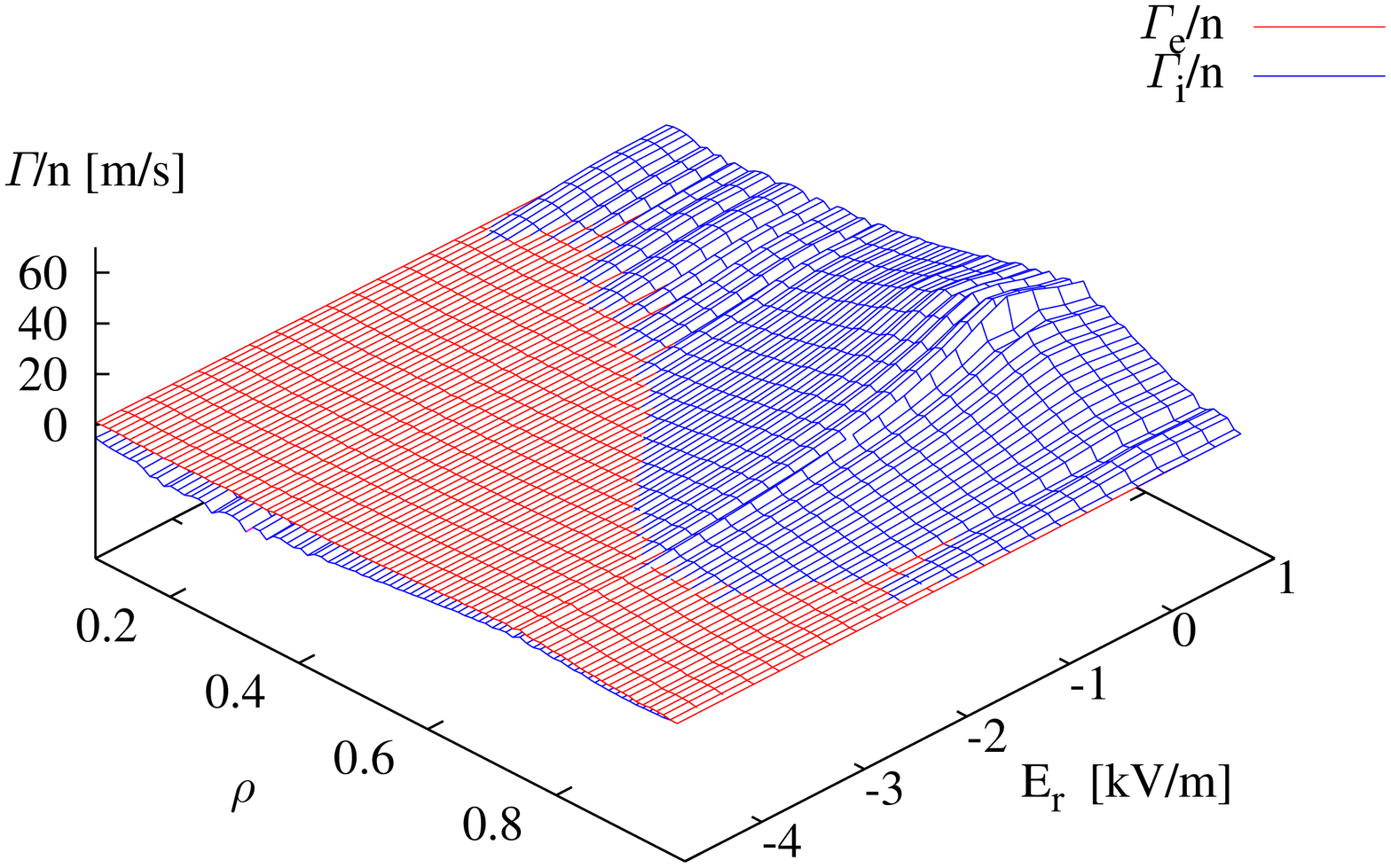}
\end{center}
%\caption{Radial profile of the solution of the ambipolarity equation for the NBI plasmas.}
\caption{Radial profile of the solution of the ambipolarity equation for the NBI plasma.}
\label{FIG_AMB3DNBI}
\end{figure}

We calculate the ambipolar electric field for two regimes of TJ-II: a
low-density ECH plasma
(Figs.~\ref{FIG_PROFILES_LD},~\ref{FIG_THERMALLD},~\ref{FIG_FLUXES_LD},~\ref{FIG_AMB2D}
and~\ref{FIG_AMB3DECH}) a high-density NBI plasma
(Figs.~\ref{FIG_PROFILES_HD},~\ref{FIG_THERMALHD},~\ref{FIG_FLUXES_HD}
and~\ref{FIG_AMB3DNBI}). Since the walls are coated with
lithium~\cite{tabares2008lithium}, TJ-II plasmas have low impurity
content, and the effective charge may be taken to be
$Z_\mathrm{eff}\!=\!1\,$ everywhere.

%First of all, the three plasmas show some common features: near the
%edge, both the temperatures and densities drop to zero in a way that
%the collisionality rises some orders of magnitude. As a result of
%this, the thermal coefficients will be very small and the neoclassical
%transport will be negligible: in this region, transport will be
%completely anomalous. Therefore, the calculated $E_\mathrm{r}$ may be
%relevant only if the dominant turbulence is electrostatic, and hence
%automatically ambipolar; the neoclassical radial fluxes are shown for
%the sake of completeness.

First of all, the two plasmas show some common features: near the
edge, both the temperatures and densities drop to zero in a way that
the collisionality rises some orders of magnitude. As a result of
this, the thermal coefficients will be very small and the neoclassical
transport will be negligible: in this region, transport will be
completely anomalous. Therefore, the calculated $E_\mathrm{r}$ may be
relevant only if the dominant turbulence is electrostatic, and hence
automatically ambipolar; the neoclassical radial fluxes are shown for
the sake of completeness.

For electrons, one has almost everywhere
$L^\mathrm{e}_{22}\!\gg\!L^\mathrm{e}_{21}\!\gg\!L^\mathrm{e}_{11}\,$,
so the temperature gradient acts usually as the main drive for both
the radial particle and energy flux. For ions, the three coefficients
$L^\mathrm{i}_{\mathrm{j}\mathrm{k}}$ are somewhat closer, and the ion
temperature is rather flat, so also the density gradient and the
radial electric field are responsible for the radial fluxes. \\

The low-density plasma has a hollow density profile and a peaked
electron temperature profile, with a central value around 1$\,$keV,
see Fig.~\ref{FIG_PROFILES_LD}. In these conditions, the electrons are
in the {\em lmfp} regime, except near the edge, where the temperature
drops. The ion temperature is lower, about 100$\,$eV, and therefore
the ions are in the plateau regime.

The thermal transport coefficients are larger for the electrons for
$\rho\!<0.6$, see Fig.~\ref{FIG_THERMALLD}, and so is the temperature
gradient. This yields a positive ambipolar radial electric field in
Fig.~\ref{FIG_FLUXES_LD}. Fig.~\ref{FIG_AMB2D} shows the solution of
the ambipolar equation at $\rho\!=\!0.22$ (top left) and
$\rho\!=\!0.61$ (top right). At both positions, the radial electron
flux, which is larger, is reduced by a non-zero electric field via the
dependence $L^\mathrm{e}_{\mathrm{j}\mathrm{k}}(E_\mathrm{r})$. The
ion radial flux is driven by the convective term
$E_\mathrm{r}/T_\mathrm{i}$ (although at $\rho\!=\!0.61$, the effect
of a radial electric field reducing the orbit width is clearly
visible). At $\rho\!=\!0.22$, the poloidal resonance may be playing a
role in $\Gamma_\mathrm{i}$ for $E_\mathrm{r}\sim 1\,kV/m$: the
diffusion coefficient has a peak for the value of $E_\mathrm{r}$ such
that the poloidal $E\times B$ drift and the poloidal parallel velocity
cancel out, and then decreases. At outer positions, the electron
collisionality is higher and the density and ion temperature gradients
rise. In these conditions, the ion particle flux is larger than the
electron flux for $\rho\!>\!0.61$. Around $\rho\!=\!0.65$ (bottom left
of Fig.~\ref{FIG_AMB2D}), two stable solutions for the ambipolar
condition appear: as in inner positions, a positive electric field
(electron root) which drives convectively the ions. But also a
negative electric field (ion root) is able to bring ion transport to
the electron level, via reduction of the radial excursions of trapped
ions. At these conditions, electrons are the {\em rate-controlling}
species (i.e., the ion radial flux is reduced to the electron level),
and the particle flux is driven by the density and electron
temperature gradients. Finally, closer to the edge, the electron
transport drops, and only a negative electric field is able to restore
ambipolarity. Bottom right of Fig.~\ref{FIG_AMB2D} shows this
happening at $\rho\!=\!0.74$. The radial dependence of the ambipolar
equation has been summarized in Fig.~\ref{FIG_AMB3DECH}, where we show
the radial particle fluxes as a function of $\rho$ and
$E_\mathrm{r}$. The intersections of
$\Gamma_\mathrm{i}(\rho,E_\mathrm{r})$ and
$\Gamma_\mathrm{e}(\rho,E_\mathrm{r})$ give the profile of the radial
electric field $E_\mathrm{r}(\rho)$.

This general behaviour is consistent with that obtained in
calculations for similar magnetic configurations of
TJ-II~\cite{tribaldos2001nctj2,turkin2011predictive} and to the
experimental data from HIBP~\cite{chmyga2002hibp,melnikov2007hibp}.

For a wide range of radial positions, $0.3\!<\!\rho\!<0.7$, the radial
fluxes are very slowly decreasing (almost constant) functions of
$\rho$. This happens because the ions, which are the rate controlling
species, are driven by a radial electric field with low shear,
$E_\mathrm{r}\sim\mathrm{d}T_\mathrm{e}/\mathrm{d}r$. A jump to higher
fluxes may be observed for $\rho\!>\!0.7$ in Fig.~\ref{FIG_FLUXES_LD},
and thus a minimum of the radial fluxes at $\rho\!=\!0.7$, the point
of maximum density and temperature gradients. The jump can also be
seen in the thermal coefficients, see Fig.~\ref{FIG_THERMALLD}: it
corresponds to the change of electron to ion root and it reflects the
difference in the drift-orbit size for the two different absolute values of
the radial electric field. The jump is thus larger for the ions, which
is consistent with the data shown in Fig.~\ref{FIG_AMB2D} (bottom
left). If $E_\mathrm{r}$ were obtained by solving a diffusion
equation, the transition from electron to ion root would be
smoother. Yet, the results of the calculations in
Ref.~\cite{turkin2011predictive} suggest that the minima might remain,
both in the particle flux and in the energy flux; the formation a
particle transport barrier at $\rho\!=\!0.7$ has been observed in {\em
low-density transitions} in ECH plasmas of TJ-II (see
Ref.~\cite{milligen2011transitions} and references therein) but no
simultaneous accumulation of energy has been detected.

During this low-density transition, a double poloidal-velocity shear
layer has been measured near the
edge~\cite{estrada2009transitions,milligen2011islands}. Although the
formation of this layer is caused by anomalous transport, once the
profiles are set the static electric field may be discussed in terms
of neoclassical fluxes: the electron temperature (and thus the thermal
transport coefficients) drops much faster than the ion temperature in
the range $0.7\!<\!\rho\!<\!0.9$, and this requires a negative shear
to maintain ambipolarity; the opposite happens for $\rho\!>\!0.9$,
hence the positive shear in our calculation.  \\

For NBI plasmas, the electron temperature is flatter and lower, since
the plasma has reached the ECH-cutoff density. The ion temperature is
still flat, and slightly higher due to the NBI heating. The lithium
wall coating has allowed~\cite{sanchez2009transitions} transitions
from plasmas of medium density ($\langle n\rangle\sim 2\times
10^{19}m^{-3}$) to plasmas of relatively high density ($\langle
n\rangle\sim 5\times 10^{19}m^{-3}$). The studied plasmas are shown in
Fig.~\ref{FIG_PROFILES_HD}: both the ions and the electrons are in the
plateau regime, the ions being slightly more collisional. In these
conditions, the thermal transport coefficients shown in
Fig.~\ref{FIG_THERMALHD} are larger for the ions, whose drift-orbit
size is much larger. Since the electron temperature is rather flat
except near $\rho\!=\!0.7$, the plasma is in the ion root everywhere,
see Fig.~\ref{FIG_FLUXES_HD}. The solution of the ambipolar equation
for these plasmas is shown in Fig.~\ref{FIG_AMB3DNBI}. The situation
is similar to that of the ECH plasma near the edge: the radial ion
flux, being larger, is reduced via $L^\mathrm{i}_{12}(E_\mathrm{r})$
to the electron level, which is in turn determined by the density and
electron temperature gradients. The radial fluxes are maximum where
the density and ion temperature gradients are larger.

The ambipolar $E_\mathrm{r}$ for the is qualitatively similar to that
obtained for a similar magnetic configurations (and lower density) in
Ref.~\cite{zurro2006rotation} and to HIBP
measurements~\cite{melnikov2007hibp}.

At TJ-II, the ion temperature is usually measured by a charge-exchange
neutral particle analyzer~\cite{fontdecaba2004cxnpa} and the profiles,
obtained on a shot-to-shot basis, are not always compatible with the
ones deduced from spectroscopy
measurements~\cite{zurro2006rotation}. It is therefore meaningful to
allow for variations in the ion temperature profile in the
neoclassical transport calculations. If $T_\mathrm{i}$ were slightly higher and
more peaked than the $T_\mathrm{i}$ employed so far, no qualitative
effects would be expected in ECH plasmas, since $E_\mathrm{r}$ is
determined by the electron temperature. The ambipolar flux, being the
ions the rate-controlling species, would probably be slightly
larger. For the NBI plasmas, a more negative electric field is
expected, maybe slightly peaked near the core region, with no major
changes in the ambipolar flux.

%\begin{figure}
%\begin{center}
%\subfigure{\includegraphics[angle=270,scale=0.25]{profiles_md.ps}}
%\subfigure{\includegraphics[angle=270,scale=0.25]{conlen_md.ps}}
%\end{center}
%\caption{Plasma profiles for the medium-density plasma: density and temperatures (left); collisionalities (right).}
%\label{FIG_PROFILES_MD}
%\end{figure}

%\begin{figure}
%\begin{center}
%\subfigure{\includegraphics[angle=270,scale=0.25]{Ljki_md.ps}}
%\subfigure{\includegraphics[angle=270,scale=0.25]{Ljke_md.ps}}
%\end{center}
%\caption{Thermal transport coefficients for the medium-density plasma.}
%\label{FIG_THERMALMD}
%\end{figure}

%\begin{figure}
%\begin{center}
%\subfigure{\includegraphics[angle=270,scale=0.25]{er_md.ps}}
%\subfigure{\includegraphics[angle=270,scale=0.25]{vrQ_md.ps}}
%\end{center}
%\caption{Radial balance for the medium-density plasma: ambipolar radial electric field (left); ambipolar particle and energy fluxes (right).}
%\label{FIG_FLUXES_MD}
%\end{figure}

\begin{table}
\begin{center}
\begin{tabular}{ccccc}
 Configuration name & $\langle B(T)\rangle_{vol}$ & V($m^3$) & $\iota(0)$ & $\iota(a)$
 \\\hline\hline
100\_32\_60 & 1.087 & 0.934 & -1.423 & -1.517 \\
100\_38\_62 & 0.971 & 1.031 & -1.492 & -1.593 \\
100\_40\_63 & 0.960 & 1.043 & -1.510 & -1.609 \\
100\_42\_63 & 0.931 & 1.079 & -1.534 & -1.630 \\
100\_44\_64 & 0.962 & 1.098 & -1.551 & -1.650 \\
100\_46\_65 & 0.903 & 1.092 & -1.575 & -1.676 \\
100\_50\_65 & 0.962 & 1.082 & -1.614 & -1.704 \\
100\_55\_67 & 0.964 & 1.073 & -1.659 & -1.739 \\
\hline
\end{tabular}
\end{center}
\caption{Main parameters of the configuration scan.}
\label{TAB_CONF}
\end{table}

\begin{figure}
\begin{center}
\subfigure{\includegraphics[angle=270,scale=0.25]{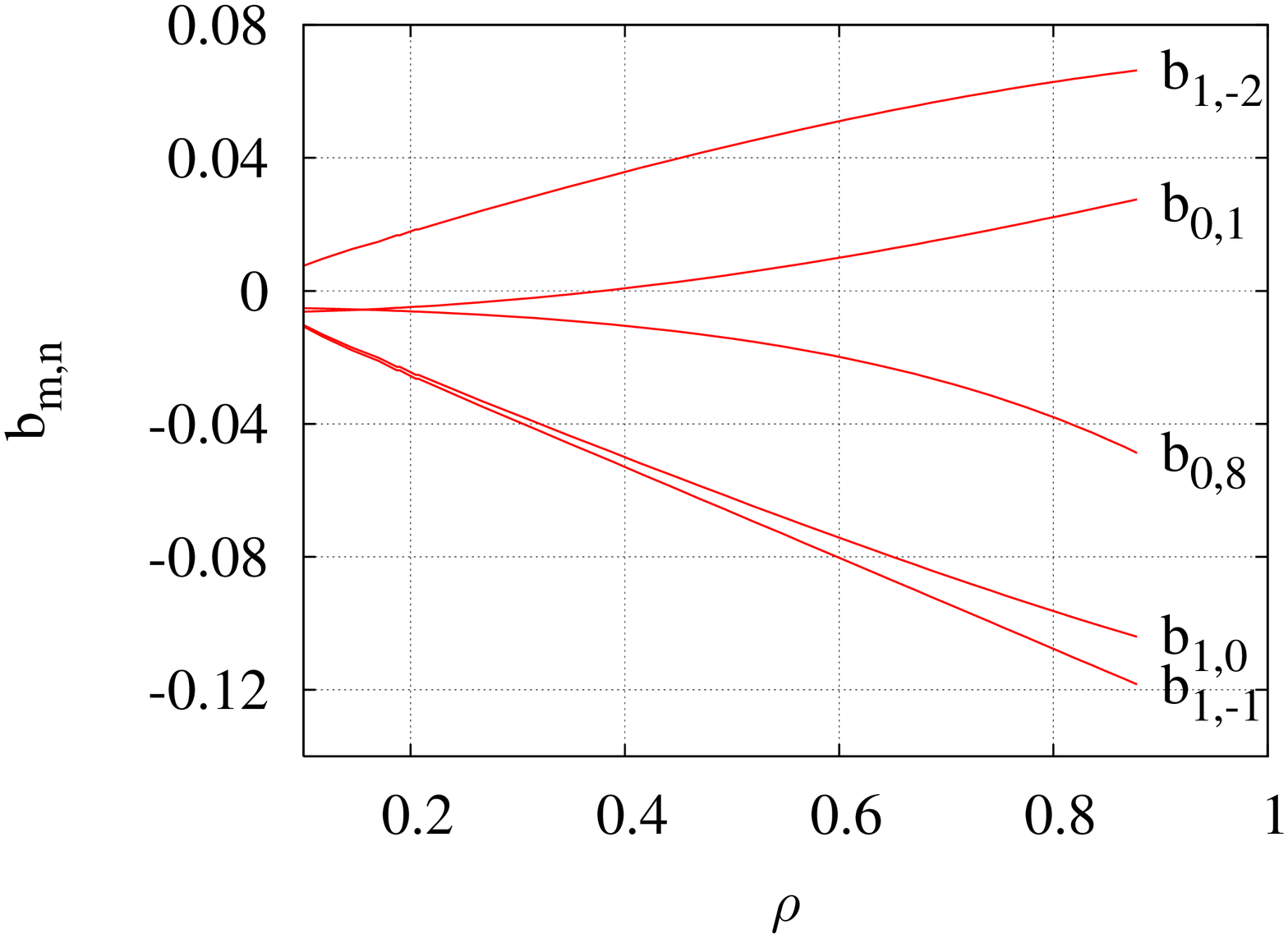}}
\subfigure{\includegraphics[angle=270,scale=0.25]{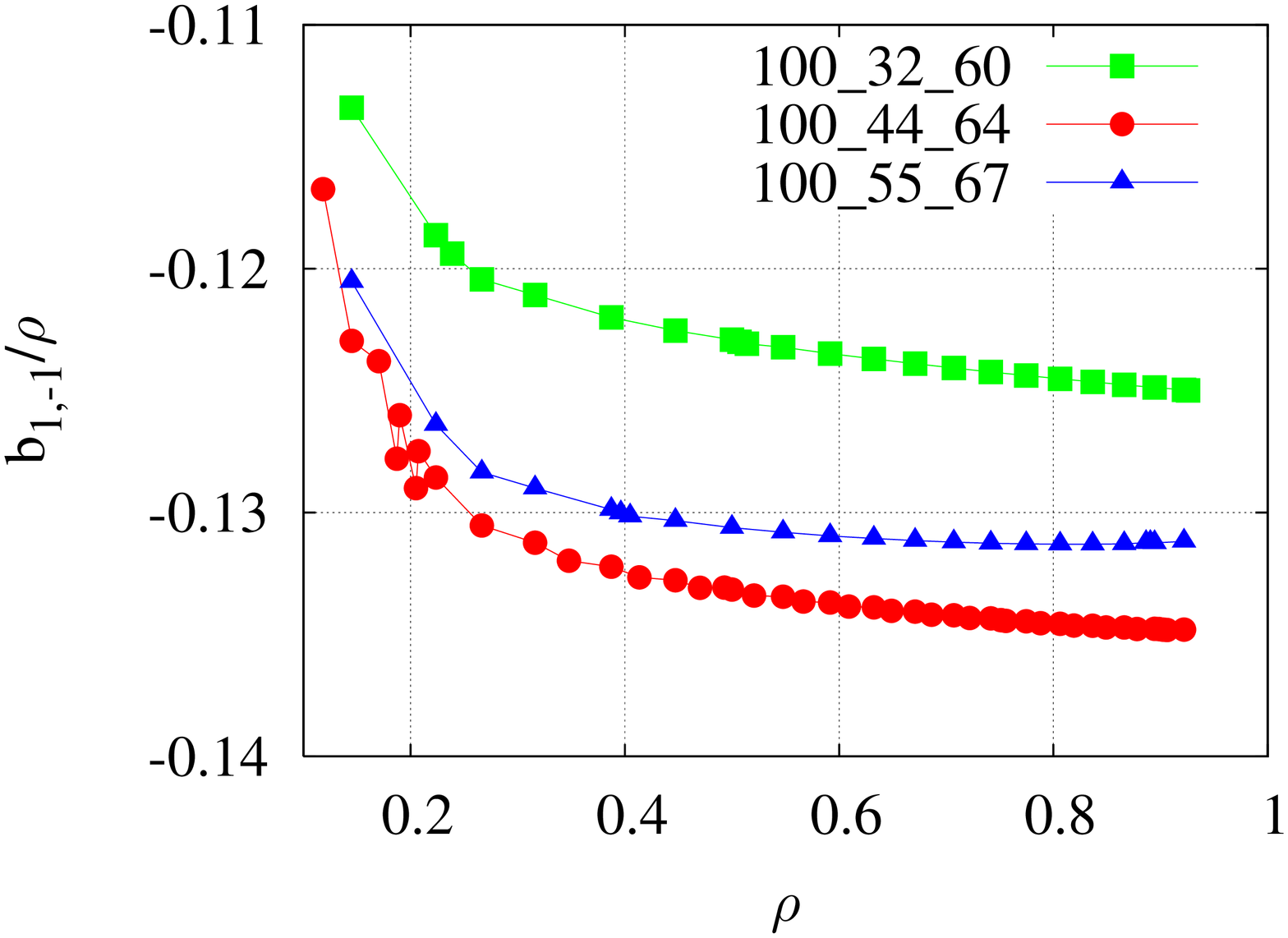}}
\subfigure{\includegraphics[angle=270,scale=0.25]{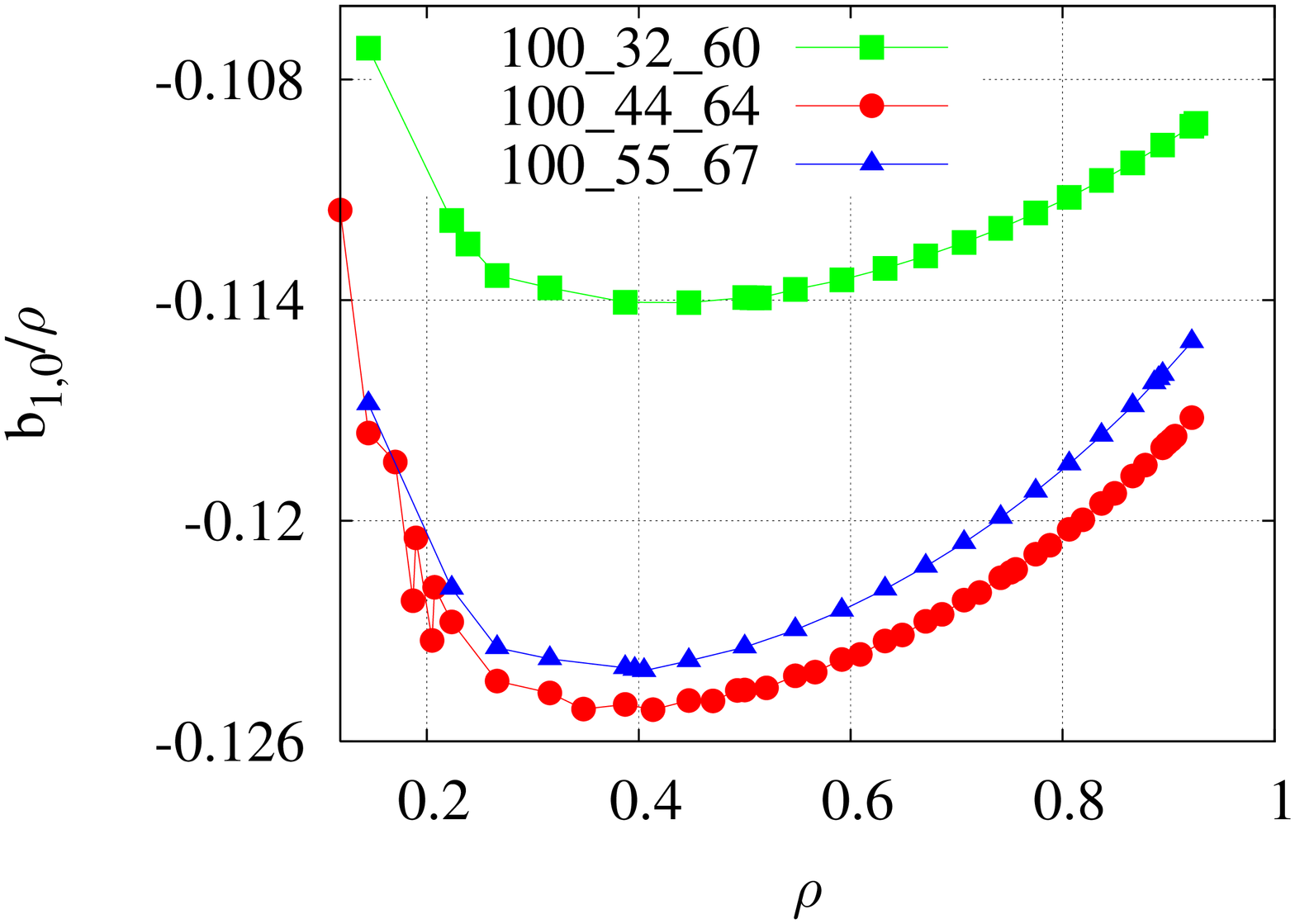}}
\subfigure{\includegraphics[angle=270,scale=0.25]{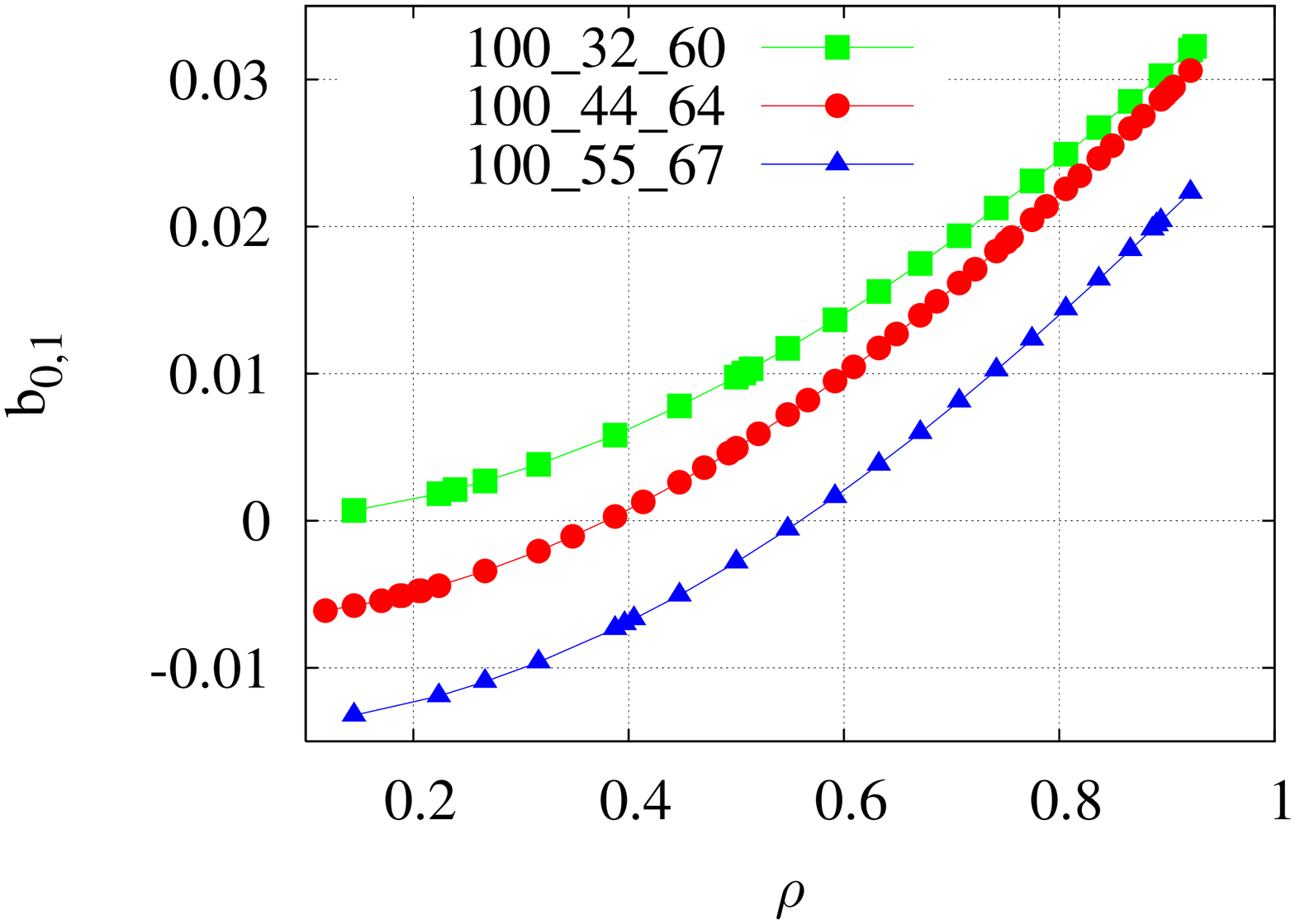}}
\end{center}
\caption{Main Fourier components of the 100\_44\_64 configuration (top
left); helical curvature (top right), toroidal curvature (bottom left)
and toroidal mirror (bottom right) for the configurations 100\_32\_60,
100\_44\_64, 100\_55\_67.}
\label{FIG_FOURIER}
\end{figure}

\begin{figure}
\begin{center}
\includegraphics[angle=270,width=1\columnwidth]{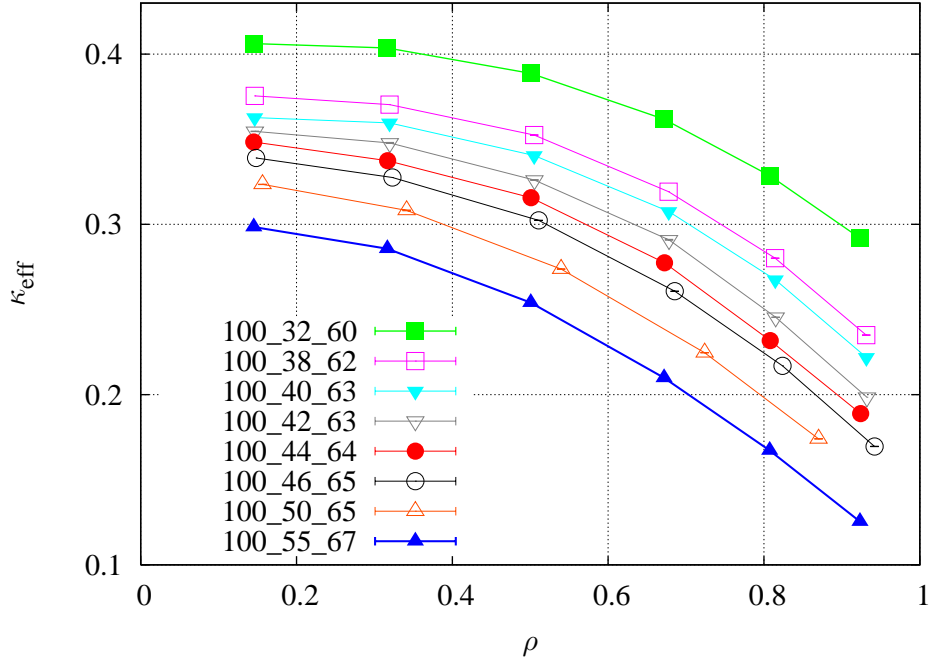}
\end{center}
\caption{Profile of the effective curvature for all the configurations of the study.}
\label{FIG_KEFF}
\end{figure}

\begin{figure}
\begin{center}
\includegraphics[angle=270,width=1\columnwidth]{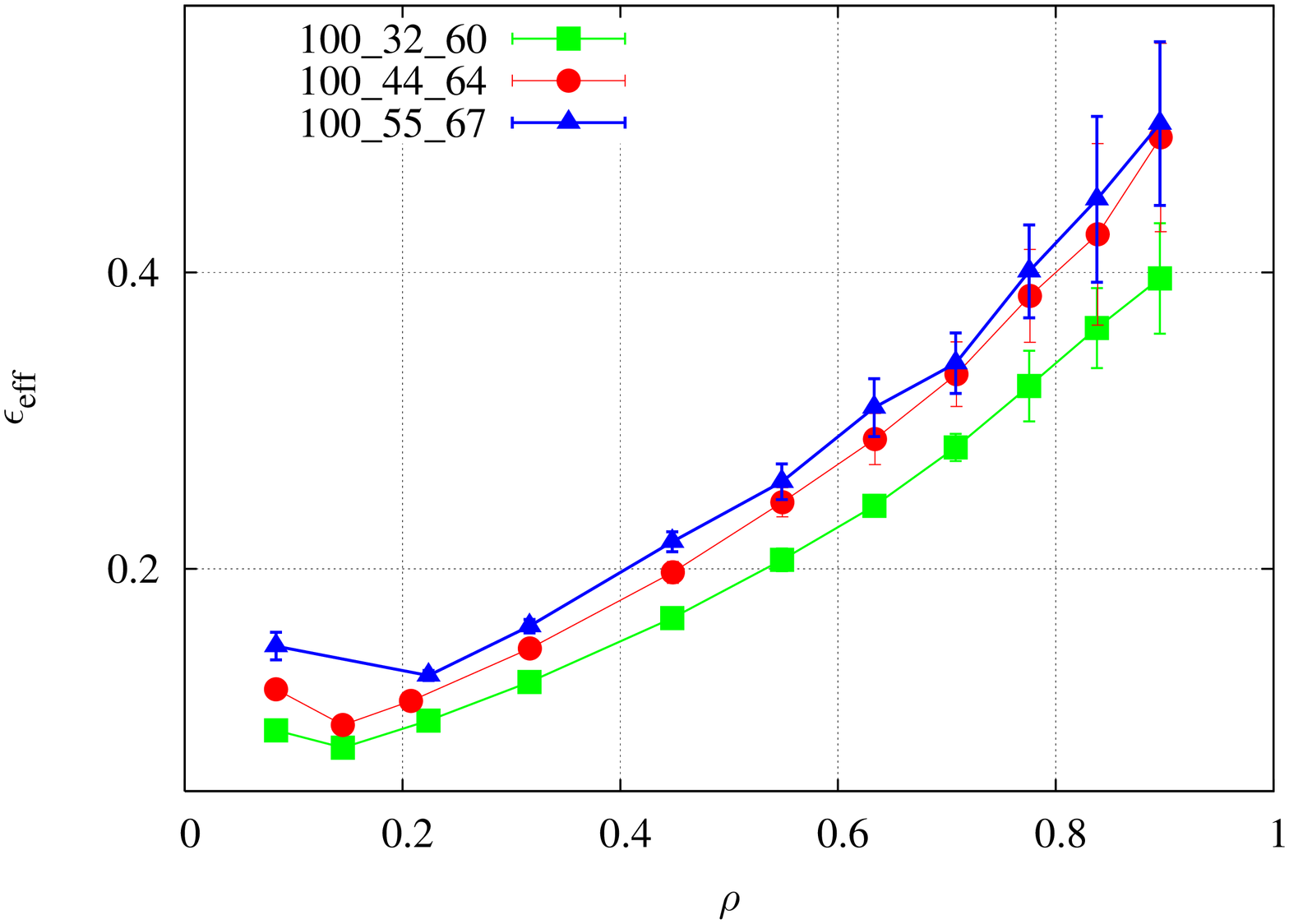}
\end{center}
\caption{Profile of the effective ripple for the configurations
100\_32\_60, 100\_44\_64, 100\_55\_67.}
\label{FIG_EPSEFF}
\end{figure}

\subsection{Configuration dependence of the radial fluxes}\label{SEC_CONFS}

Finally, we make a scan in a relevant part the configuration space of
TJ-II. We study eight magnetic configurations usually explored in
regular operation of TJ-II. Their main global parameters are shown in
Table~\ref{TAB_CONF}. The rotational transform at the magnetic axis
varies about a 20\% during the scan (and so does at the edge, since
the profile shape is kept unchanged). Since the volume is
approximately constant as well and the main Fourier terms in the
description of the magnetic field do not change too much, no large
differences in the neoclassical radial transport are
expected~\cite{solano1988tj-ii}. Indeed the qualitative behaviour is
identical to that of Fig.~\ref{FIG_MONO}. We thus make a discussion on
the radial profile of several quantities that parametrize the
dependence on $\nu^*$ of the monoenergetic coefficient for the
100\_44\_64 configuration and the others. \\

The two main contributions to the radial diffusion should come from
the helical ($b_{1,-1}$) and toroidal ($b_{1,0}$) curvatures and the
toroidal mirror ($b_{0,1}$) in the Fourier decomposition of $B$ given
by:
\begin{equation}
B(\rho,\phi,\theta)/B_0(\rho)=\sum_{n=-\infty}^{\infty}\sum_{m=0}^{\infty}b_{m,n}(\rho)cos(m\theta-Nn\phi)\,,
\label{EQ_FOURIER}
\end{equation}
where $\phi$ and $\theta$ are Boozer coordinates. They are shown in
Fig.~\ref{FIG_FOURIER} for three of the configurations of the scan:
100\_44\_64, 100\_32\_60, 100\_55\_67 (the center and the extremes of
the scan). \\

In the PS regime, for $\Omega\!=\!0\,$, we have
$D^*_{11}\!\propto\!\kappa^{-1}\nu^*\,$~\cite{igitkhanov2006impurity}. Here,
$\kappa\propto(b_{1,0})^{-2}$ is called {\em toroidal curvature}. The
small correction for large $\Omega$ depends weakly on $\iota$ and $R$.

According to Fig.~\ref{FIG_FOURIER}, we can expect the collisional
transport of these configurations to be quite similar. Nevertheless,
modulation of the toroidal mirror term $b_{0,1}$ allows for
optimization of radial transport in elongated configurations (see
e.g. Ref~\cite{beidler2011ICNTS}). In Ref.~\cite{aizawa2000curvature},
an {\em effective toroidal curvature} was defined, including $b_{1,0}$
and $b_{0,1}$, in order to account for this effect. The latter Fourier
term has a large relative (although not absolute) variation in this
configuration scan~\cite{velasco2011bootstrap}. From the slope of
$D^*_{11}$ for high $\nu^*$ in Fig.~\ref{FIG_MONO} one can calculate
an {\em effective curvature} $\kappa_\mathrm{eff}$ defined by:
\begin{equation}
D^*_{11}(\nu^*\to\infty,\Omega=0)=\frac{32}{3\pi}\frac{\nu^*}{\kappa_\mathrm{eff}}\,.
\end{equation}
Therefore, from the slope of $D^*_{11}$ for high $\nu^*$ in
Fig.~\ref{FIG_MONO} one can extract the local value of
$\kappa_\mathrm{eff}$. The profile of this quantity is shown in
Fig.~\ref{FIG_KEFF}. There is a $\iota^{-2}$ scaling (see
Table~\ref{TAB_CONF}) at every radial position in our set of
configurations, so we focus, from now on, on the extremes of the scan.
Since $D_{11}^\mathrm{p}\!\propto\!(\iota B_0^2)^{-1}$ and
$\nu^*\!\propto\!\iota^{-1},$ the contribution of the PS regime to the
thermal transport coefficient is
$L_{11}\!\propto\!(\kappa_\mathrm{eff}\iota^2 B_0^2)^{-1}$. These
results altogether yield a neoclassical transport in the configuration
100\_32\_60 reduced with respect to the others (100\_44\_64 and
100\_55\_65) via the trivial scaling $L_{11}\!\propto\!B_0^{-2}\,$. \\

The results are different in the {\em lmfp} regime. In the limit case
of classical stellarator (i.e., if only $b_{1,0}$ and $b_{1,-1}$ are
non-zero in Eq.~(\ref{EQ_FOURIER})), we have
$D^*_{11}\!\propto\!(b_{1,-1})^{3/2}/\nu^*\,$ for
$\Omega\!=\!0$~\cite{galeev1979theory}. If more Fourier terms are
non-zero, particle-trapping in local minima of the magnetic field
leads to enhanced radial transport. However, one can still describe
the radial diffusion in the {\em lmfp} regime in terms of an {\em
effective helical ripple}
$\varepsilon_\mathrm{eff}\,$\cite{dommaschk1984ripple,beidler1994ripple,beidler2011ICNTS}. This
quantity is defined by:
\begin{equation}
D^*_{11}(\nu^*\to 0,\Omega=0)=\left(\frac{4}{3\pi}\right)^2\frac{(2\varepsilon_\mathrm{eff})^{3/2}}{\nu^*}\,,
\end{equation}
obtained from data such as Fig.~\ref{FIG_MONO}. It contains
information of the helical ripple $b_{1,-1}$, as well as of all the
other terms in Eq.~(\ref{EQ_FOURIER}). In Fig.~\ref{FIG_EPSEFF} we
show the radial profile of the effective ripple for three of the
configurations: 100\_32\_60, 100\_44\_64, 100\_55\_67. The
configuration 100\_32\_60 has a smaller effective ripple, while those
of the 100\_44\_64 and 100\_55\_67 configurations are quite
close. Since, for the {\em lmfp} regime, one has
$L_{11}\!\propto\!\varepsilon_\mathrm{eff}^{3/2}B_0^{-2}\,$, the
latter configurations will have similar radial neoclassical
transport. Configuration 100\_32\_60 will have considerably smaller
transport, and the reduction will be larger than the
$L_{11}\!\propto\!B_0^{-2}\,$ of the PS regime. \\

\begin{figure}
\begin{center}
\subfigure{\includegraphics[angle=270,scale=0.25]{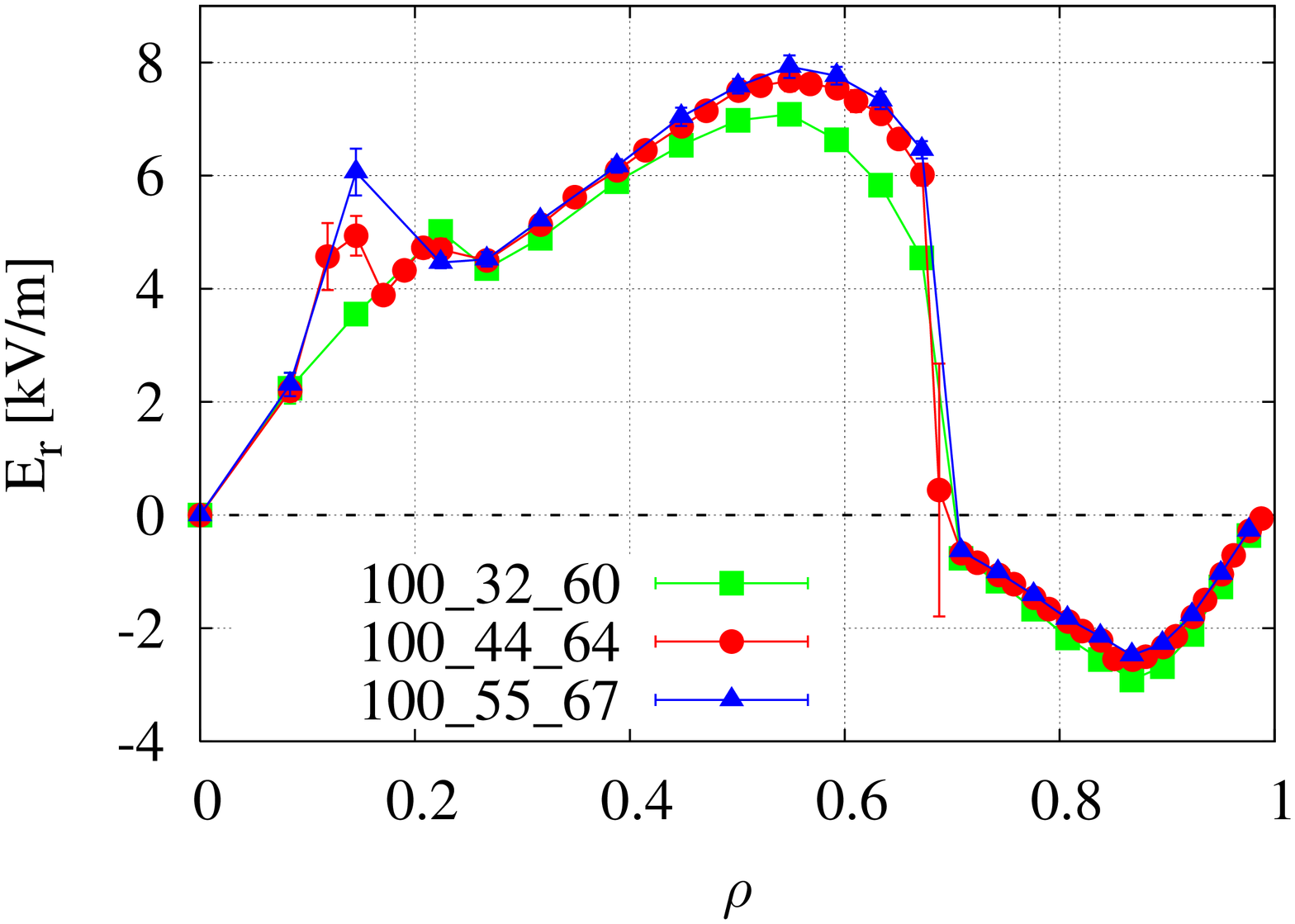}}
\subfigure{\includegraphics[angle=270,scale=0.25]{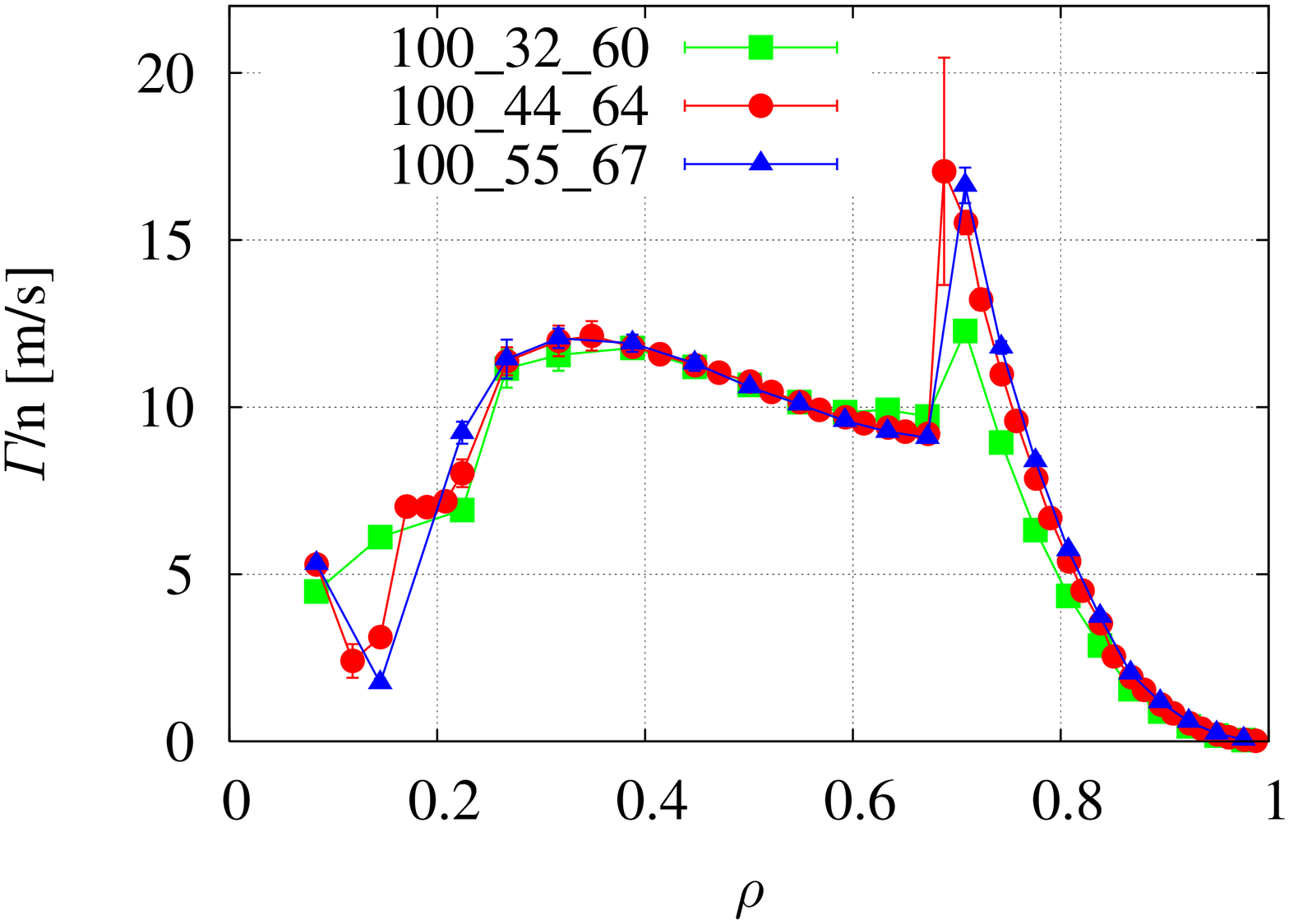}}
\end{center}
\caption{Ambipolar radial electric field (left) and radial particle
fluxes (right) for configurations 100\_32\_60, 100\_44\_64,
100\_55\_67 and the profiles of Fig.~\ref{FIG_PROFILES_LD}.}
\label{FIG_ERCONF_ECH}
\end{figure}

\begin{figure}
\begin{center}
\subfigure{\includegraphics[angle=270,scale=0.25]{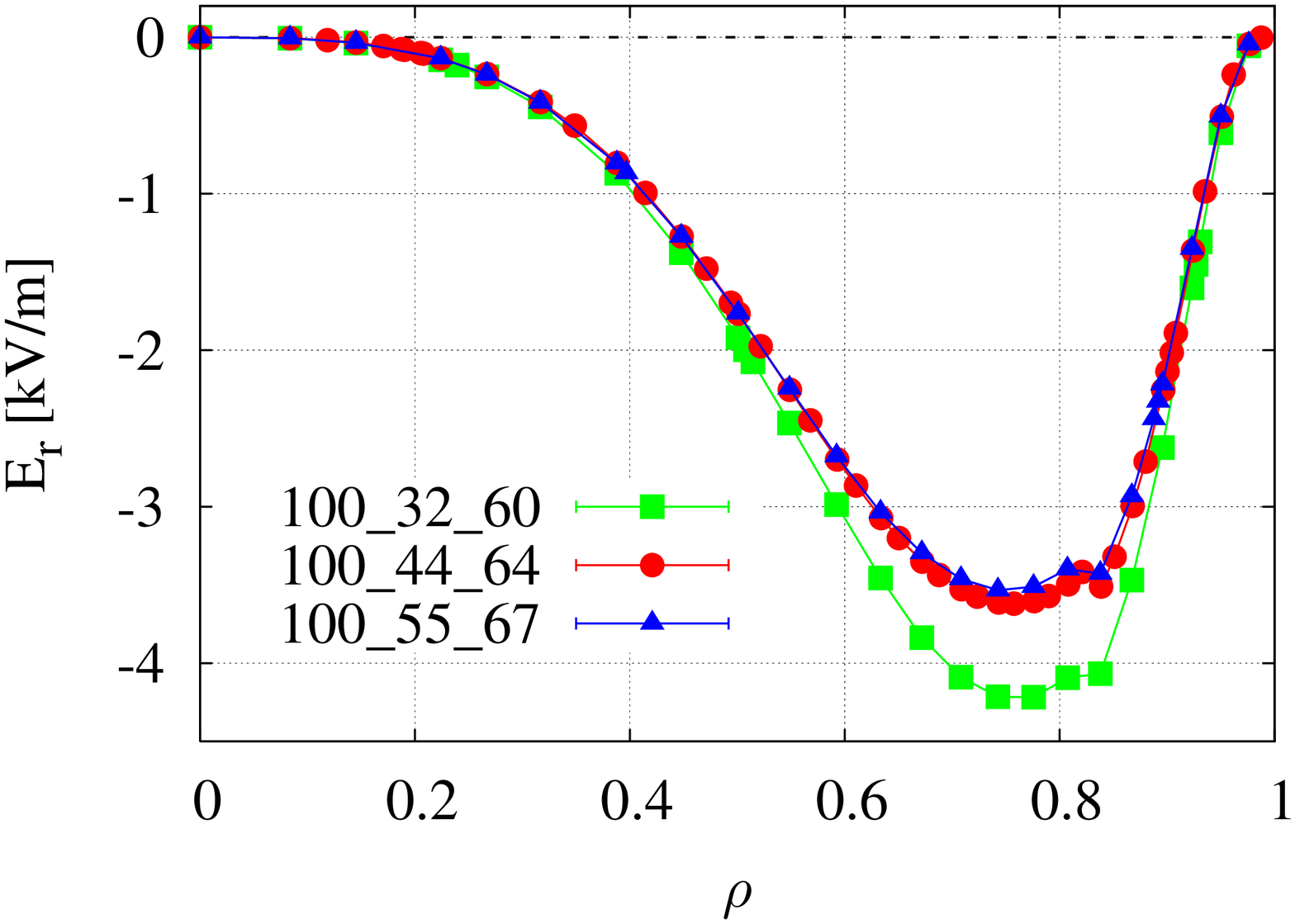}}
\subfigure{\includegraphics[angle=270,scale=0.25]{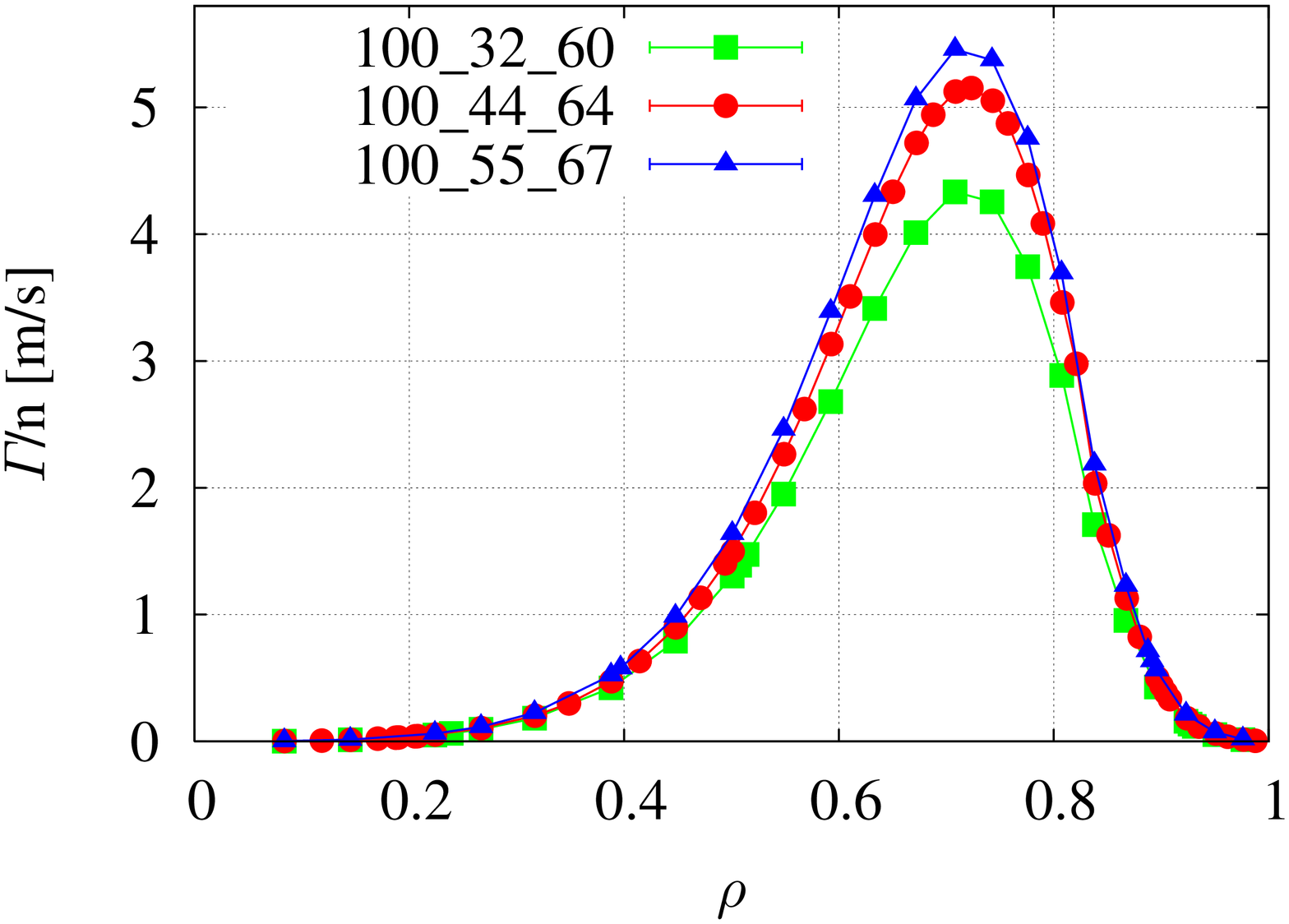}}
\end{center}
\caption{Ambipolar radial electric field (left) and radial particle
fluxes (right) for configurations 100\_32\_60, 100\_44\_64,
100\_55\_67 and the profiles of Fig.~\ref{FIG_PROFILES_HD}.}
\label{FIG_ERCONF_NBI}
\end{figure}

These results altogether lead to a reduced radial neoclassical
transport for the 100\_32\_60 configuration with respect to the
100\_44\_64, 100\_55\_67 configurations. The reduction is larger for
the electrons (which have a larger contribution of the {\em lmfp}
regime) than for the ions (which are more collisional). Consequently,
this will affect, via the ambipolar condition, the neoclassical radial
electric field, which will be smaller. This is consistent with
impurity poloidal rotation measurements for ECH plasmas in
Ref.~\cite{zurro2006rotation}. The results are shown in
Fig.~\ref{FIG_ERCONF_ECH} for the profiles of the ECH plasma, see
Fig.~\ref{FIG_PROFILES_LD}: the differences in $E_\mathrm{r}$ are
larger for $\rho\!<\!0.7$, where the collisionality is lower. The
radial particle fluxes are mainly reduced for $\rho\!>\!0.7$, where
the ion-root is realized and thus the electrons are the
rate-controlling species. The results for the NBI plasmas, shown in
Fig.~\ref{FIG_ERCONF_NBI}, are similar to that of the ECH plasma for
$\rho\!>\!0.7$: a small reduction of the electric field (which now
becomes more negative) together with a reduction of the ambipolar
particle flux.

\section{Conclusions}\label{SEC_CONCLUSIONS}

%We have presented calculations of the radial diffusion monoenergetic
%transport coefficient for several configurations of TJ-II. By
%convolution of these coefficients, we have calculated the ambipolar
%radial electric field and fluxes for three different plasmas
%corresponding to the 100\_44\_64 configuration. The convolution
%included data with large error bars due to the poor convergence of
%DKES in the {\em lmfp} regime, but a Monte Carlo method for error
%propagation has allowed us to show that the results are accurate even
%for the ECH plasma.

%The results for the ECH plasma and the medium-density NBI plasmas are
%similar to the ones obtained in previous works. No such calculations
%exist for high-density NBI plasmas, but the results shown here stay in
%qualitative agreement with measurements. Future work includes
%comparison with HIBP and CXRS measurements for this kind of
%discharges.

We have presented calculations of the radial diffusion monoenergetic
transport coefficient for several configurations of TJ-II. By
convolution of these coefficients, we have calculated the ambipolar
radial electric field and the fluxes for two different plasmas
corresponding to the 100\_44\_64 configuration. The convolution
included data with large error bars due to the poor convergence of
DKES in the {\em lmfp} regime, but a Monte Carlo method for error
propagation has allowed us to show that the results are accurate even
for the ECH plasma.

The results for the ECH plasma are similar to the ones obtained in
previous works. No such calculations exist for high-density NBI
plasmas, but the some of the results shown here stay in qualitative
agreement with the experiment. Future work includes comparison with
HIBP and CXRS measurements for this kind of discharges.

Small quantitative and no qualitative differences have been found
between configurations, since the Fourier spectra are very
similar. The results predict that configurations with reduced $\iota$
lead, for the same plasma profiles, to slightly smaller radial
electric field and to slightly improved particle confinement.

The results shown here extend the knowledge of neoclassical transport
and radial electric field to regimes and configurations not explored
previously at TJ-II.

\section{Acknowledgments}

The authors are grateful to D. Spong and S.P. Hirshman for the DKES
and VMEC codes. Previous discussions with H. Maa{\ss}berg,
C.D. Beidler and A. L\'opez-Fraguas, were very useful. Conversations
with B Ph. Van Milligen, B. Zurro and I. Calvo improved the quality of
the manuscript. This work has been partially funded by the Spanish
Ministerio de Ciencia e Innovaci\'on, Spain, under Project
ENE2008-06082/FTN.

\section*{References}

\bibliographystyle{iopart-num} \bibliography{}

\end{document}